\documentclass[twocolumn,showpacs,preprintnumbers,amsmath,amssymb]{revtex4}
\usepackage{epsfig}
\usepackage{dcolumn}
\usepackage{bm}

\newcommand{\remove}[1]{}
\newcommand{\dd}{\mathrm{d}}

\def\be{\begin{equation}}
\def\ee{\end{equation}}

\newcommand{\beq}{\begin{equation}}
\newcommand{\eeq}{\end{equation}}
\newcommand{\beqa}{\begin{eqnarray}}
\newcommand{\eeqa}{\end{eqnarray}}

\renewcommand{\pl}{\partial}

\newcommand{\vv}{{\bf v}}

\newcommand{\vx}{{\bf x}}
\newcommand{\vk}{{\bf k}}
\newcommand{\vp}{{\bf p}}

\newcommand{\tdelta}{{\tilde{\delta}}}

\newcommand{\ttheta}{{\tilde{\theta}}}

\newcommand{\cF}{{\cal F}}
\newcommand{\cG}{{\cal G}}
\newcommand{\cH}{{\cal H}}

\newcommand{\cP}{{\cal P}}

\newcommand{\rhob}{\overline{\rho}}
\newcommand{\Om}{\Omega_{\rm m}}

\newcommand{\bea}{\begin{array}}
\newcommand{\ea}{\end{array}}

\begin{document}

\title{Impact of a Warm Dark Matter late-time velocity dispersion on large-scale structures}

\author{Patrick Valageas}
\affiliation{Institut de Physique Th\'eorique,\\
CEA, IPhT, F-91191 Gif-sur-Yvette, C\'edex, France\\
CNRS, URA 2306, F-91191 Gif-sur-Yvette, C\'edex, France}
\vspace{.2 cm}

\date{\today}
\vspace{.2 cm}

\begin{abstract}
We investigate whether the late-time (at $z\leq 100$) velocity dispersion expected in
Warm Dark Matter scenarios could have some effect on the cosmic web (i.e., outside
of virialized halos).
We consider effective hydrodynamical equations, with a pressurelike term that agrees
at the linear level with the analysis of the Vlasov equation. Then, using analytical 
methods, based on perturbative expansions and the spherical dynamics,
we investigate the impact of this term for a $1$keV dark matter particle.
We find that the late-time velocity dispersion has a negligible effect
on the power spectrum on perturbative scales and on the halo mass function.
However, it has a significant impact on the probability distribution function of the density
contrast at $z \sim 3$ on scales smaller than $0.1 h^{-1}$Mpc, which correspond to
Lyman-$\alpha$ clouds. Finally, we note that numerical simulations should start
at $z_i\geq 100$ rather than $z_i \leq 50$ to avoid underestimating gravitational clustering
at low redshifts.

\keywords{Cosmology \and large scale structure of the Universe}
\end{abstract}

\pacs{98.80.-k,95.35.+d} \vskip2pc

\maketitle

\section{Introduction}
\label{Introduction}

In the standard cosmological $\Lambda$ cold dark matter ($\Lambda$CDM) scenario,
most of the matter content
of the Universe is made of CDM particles, which are cold and
collisionless. This means that they have a negligible velocity dispersion during the
matter-dominated era and density fluctuations on almost all scales (except very small
scales at early times) grow through gravitational instability.
This leads to a hierarchical scenario for the formation of large-scale structures,
as the amplitude of density fluctuations at early time (e.g., at the beginning of the
matter-dominated era) is larger on smaller scales. Then, small scales turn nonlinear
first and merge to build increasingly large and massive structures as larger scales
become nonlinear in the course of time \cite{Peebles1980}.
This scenario (with an extra dark energy component or cosmological constant) is in
good agreement with a large variety of cosmological observations, such as the cosmic
microwave background (CMB) \cite{Komatsu2011} and galaxy surveys \cite{Tegmark2006}.

However, this CDM model may disagree from observations on small scales (below the
size of galaxies). Thus, CDM simulations typically predict too many
satellite galaxies around Milky-Way-sized central galaxies as compared with observations
\cite{Moore1999,Springel2008,Trujillo-Gomez2011}.
They also predict power-law density profiles, $\rho \sim r^{-1}$, in the center of virialized
halos \cite{Navarro1997},
whereas dark-matter-dominated dwarf galaxies \cite{Burkert1995} and some disk galaxies 
\cite{Salucci2000} exhibit flat density cores. This is the so-called 
``core-cusp problem'' \cite{de-Blok2010}.

One possible solution to these small-scale problems is a warm dark matter (WDM)
scenario, with dark matter particles of a mass on the order of $1$keV.
This intermediate case between the ``cold'' and ``hot'' dark matter scenarios
provides a non-negligible velocity dispersion and a significant free-streaming that erases
density fluctuations on small scales (mostly during the period where the particles
are relativistic). This helps to cure the small-scale problems of
the CDM scenario, while being
indistinguishable from CDM on large scales, which preserves its good agreement
with large-scale observations such as the CMB and galaxy surveys
\cite{Bode2001,Avila-Reese2001,Menci2012,Lovell2012}.
This favors a mass on the order of $1$keV \cite{de-Vega2010,de-Vega2012}.
For larger masses we recover the
CDM scenario and for smaller masses we recover the hot dark matter scenario, where
structure formation begins too late (in particular, this is ruled out by the Gunn-Peterson
bound \cite{Gunn1965}: quasar spectra show that the Universe must have been reionized
before $z \sim 6$, which requires galaxy formation by this time). Similar lower bounds on $m$
are also obtained from the observed velocity dispersion of dwarfs galaxies and from the
Lyman-$\alpha$ forest \cite{Viel2005,Seljak2006,Abazajian2006a,Boyarsky2009}.

We must note that these small-scale problems may also be cured by the physics of
the baryonic component, within the CDM scenario.
For instance, reionization of the intergalactic medium \cite{Bullock2000,Benson2002}
or feedback from stars and supernovae \cite{Kauffmann1993} suppress star formation
in small satellite halos. Only a small fraction of the low-mass dark
matter satellites would then shine in the sky and appear in galaxy surveys. This would
reconcile the observed abundance with the CDM prediction but there remains some
discrepancy for the shape of the satellite luminosity function \cite{Koposov2008}.
Supernovae explosions may also transform a cusp density profile
into a cored one, within small dark matter halos 
\cite{Mashchenko2006,Governato2010,de-Souza2011}.
However, it is a difficult task to check that such models can explain galaxy
properties from massive to dwarf galaxies and from $z=0$ to higher redshifts \cite{Font2011}.

Therefore, WDM scenarios remain interesting alternatives to CDM that are still being
investigated in many works.
As recalled above, a particle mass on the order of $1-10$keV is a good candidate and it
may correspond, for instance, to sterile neutrinos 
\cite{Dodelson1994,Abazajian2001,Abazajian2006,Shaposhnikov2006,Boyarsky2009a,Kusenko2009,Abazajian2012}
or to gravitinos \cite{Kawasaki1997,Gorbunov2008}.
At early times and on large scales, the formation of large-scale structures within WDM scenarios
is studied through the linearized Vlasov equation 
\cite{Boyanovsky2008,Boyanovsky2011,Boyanovsky2011a,de-Vega2012a,de-Vega2012b}.
At low redshift and on small scales, the nonlinear regime of gravitational clustering is
investigated through numerical simulations \cite{Bode2001,Schneider2011}
and halo models based on such simulations \cite{Smith2011,Dunstan2011}.
In practice, one often uses the same N-body codes as for CDM scenarios and the only
difference comes from the density power spectrum that is set at the initial redshift $z_i$
of the simulations. This means that one takes into account the high-$k$ cutoff due to
free-streaming during the relativistic era but neglects the nonzero velocity dispersion at
low redshifts, $z\leq z_i$.
This is legitimate because the relative importance of this late-time velocity dispersion
decreases with time and the main difference between the CDM and WDM scenarios
with respect to large-scale structures arises from the high-$k$ cutoff of the power spectrum
built during the relativistic regime.
Nevertheless, it would be interesting to have a quantitative check of this approximation.
This is the goal of this paper, where we obtain a quantitative estimate of the impact
of the late-time WDM velocity dispersion on the formation of large-scale structures.
We also estimate the sensitivity of the gravitational clustering measured at low redshift
on the initial redshift $z_i$ of the simulations.

Here we do not consider the inner regions of virialized halos, where the finite velocity
dispersion can have important effects because of Liouville theorem. Indeed, this implies
an upper bound on the coarse-grained phase-space distribution function
\cite{Tremaine1979}, which can lead to cored density profiles instead of cusps
\cite{Hogan2000} (but the behavior in central regions remains difficult to predict
\cite{Vinas2012}). 
In contrast, we consider the cosmic web, that is, moderate
density fluctuations or large scales, as well as the halo mass function itself.
Then, using perturbative methods or the spherical dynamics, we compare such statistics
(the power spectrum on large perturbative scales, the halo mass function, and the probability
distribution of the density contrast) between the CDM and WDM scenarios, where we
neglect or take into account the late-time WDM velocity dispersion.
To simplify the analysis and to go beyond the linear regime, we use
effective equations of motion similar to standard hydrodynamics.
They involve a simplified pressurelike term in the Euler equation, associated with the
late-time velocity dispersion, that is chosen so as to agree with results from the Vlasov
equation at linear order.
This should be sufficient for our goal, which is only to estimate the order of magnitude
of the impact of this late-time WDM velocity dispersion.
Thus, our study is complementary to Ref.\cite{Boyanovsky2011a} who investigates the effects
of the velocity dispersion at low redshift through the linearized Vlasov equation.
Our approach is not exact, because we use a fluid approximation, but it allows us to
consider nonlinear density fluctuations.
In particular, our goal is not to study in accurate details a specific WDM model but
to investigate the generic impact of a late-time velocity dispersion on the cosmic web.
After describing these effective equations of motion and our approach in
Sec.~\ref{motion}, we present our results for a $1$keV dark matter particle 
in Sec.~\ref{Results}.
We also consider the impact of the choice of initial redshift in numerical simulations
in Sec.~\ref{lower-zi} and we conclude in Sec.~\ref{Conclusion}.

\section{Equations of motion}
\label{motion}

\subsection{Effective Euler equation}
\label{Effective-Euler}

During the matter-dominated era, on scales much smaller than the Hubble radius and
for nonrelativistic particles, the evolution of the dark matter density perturbations is
governed by the nonrelativistic Boltzmann-Vlasov equation \cite{Peebles1980},
\beqa
&& \frac{\pl f}{\pl\tau} + \frac{\vp}{m a} \cdot \frac{\pl f}{\pl \vx} - m a \frac{\pl\Phi}{\pl\vx}
\cdot \frac{\pl f}{\pl \vp} = 0 ,  \label{Vlasov} \\
&& \Delta \Phi =\frac{4\pi\cG m}{a} \left( \int \dd\vp \, f - \bar{n} \right) . \label{Poisson}
\eeqa
Here $\tau$ is the conformal time, $a(t)$ the scale-factor, $\vx$ and $\vp$ are comoving 
coordinate and momentum, $m$ is the mass of the particles, $\bar{n}$ the mean dark matter 
number density, and $f(\vx,\vp,\tau)$ the phase-space distribution function.
Because the gravitational potential $\Phi$ depends on the dark matter density fluctuations
(here we neglected the contribution from baryons, or, more precisely, we neglect the
difference between baryons and dark matter), the Vlasov equation (\ref{Vlasov})
is nonlinear over $f$. Then, one usually linearizes the system (\ref{Vlasov})-(\ref{Poisson})
to study the evolution of the dark matter perturbations 
\cite{Boyanovsky2008,Boyanovsky2011}.

The phase-space distribution $f(\vx,\vp,\tau)$ includes the distribution of velocities of the
dark matter particles at any point in space. However, for observational purposes one is
mostly interested in the density and peculiar velocity fields, 
which are the lowest-order moments of $f$,
\beq
\int \dd\vp \; f = \frac{\rho}{m} , \hspace{1cm} \int \dd\vp \; \vp \; f = a \rho \vv .
\label{v-def}
\eeq
By taking successive moments over $\vp$ of the Vlasov equation (\ref{Vlasov}) one
obtains an infinite hierarchy of equations that no longer depend on the velocity coordinate
\cite{Peebles1980,Shoji2010}.
However, contrary to the case of collisional fluids, this hierarchy cannot be closed by
writing the second-order moment (the velocity dispersion) as a pressure that obeys
a definite equation of state.
In the case of cold dark matter, this hierarchy can still be closed by neglecting the velocity
dispersion (whence the name ``cold'') and one is left with only two fields, the density
$\rho(\vx,\tau)$ and the mean peculiar velocity $\vv(\vx,\tau)$. 
In the case of warm dark matter, the velocity dispersion cannot be neglected at high
redshift and one must work with the Vlasov equation.
This large velocity dispersion leads to a significant free streaming that damps the early time
density power spectrum at high $k$, as compared with CDM,
because particles can escape from small potential wells \cite{Bode2001}.
This is very efficient at early times, when the particles are still relativistic.
Therefore, the damping is greater for particles of smaller mass $m$, which become
nonrelativistic later.
Afterwards, because the typical comoving velocity dispersion decreases with time because
of the Hubble expansion, these effects become less important on large scales.
(However, because of Liouville theorem, the finite velocity dispersion, which leads to a finite 
value for the initial phase-space density $f$ --as opposed to the cold case where it would be a 
Dirac distribution-- leads to an upper bound for the coarse-grained phase-space
distribution \cite{Tremaine1979}. This can have a significant effect on the matter distribution
on small scales, within collapsed halos, even today \cite{Hogan2000}.)

To go beyond the linear regime, one usually runs numerical simulations to investigate
the formation of the cosmic web and collapsed halos.
In practice, one often uses the same codes as for CDM 
\cite{Colombi1996a,Bode2001,Schneider2011,Dunstan2011} and the only
difference is encapsulated in the initial density power spectrum set at the initial redshift $z_i$
of the simulation (i.e., the high-$k$ cutoff due to early time free streaming).
Sometimes, in order to mimic the effect of the upper bound for the coarse-grained
phase-space distribution, one adds an initial white-noise velocity component to the
particles of the simulation \cite{Colin2008,Boyarsky2009b,Viel2012}.
However, this is not truly equivalent to the full phase-space distribution $f(\vx,\vp,\tau)$
because the ``particles'' used in the simulations are actually ``macroparticles'', which
model large clumps of matter, rather than the physical dark matter particles.
 
Our goal in this paper is to estimate the impact of a late-time WDM
velocity dispersion on large-scale gravitational clustering.
As explained above, within a hydrodynamical approach, one cannot truly close the 
hierarchy of equations obeyed by the successive moments of the phase-space distribution.
However, because the fluid equations are much easier to follow beyond the linear level
than the Vlasov equation and we only look for an order-of-magnitude estimate in this
simple study, we will use a simple closure inspired from linear theory.
As shown for instance in \cite{Boyanovsky2008,Boyanovsky2011}, 
in the matter dominated era, one can derive
from the linearized Vlasov equation an evolution equation for the matter density contrast of the
form
\beq
\frac{\pl^2 \tdelta}{\pl\tau^2} + \cH \frac{\pl\tdelta}{\pl\tau}
- \left( \frac{3}{2} \Om \cH^2 - k^2 c_s^2 \right) \tdelta = S[\tdelta,\tau] ,
\label{delta-Cs}
\eeq
where $\Om(\tau)$ is the matter cosmological parameter, $\cH=\dot{a}$ is the conformal
expansion rate, and we denote with a tilde Fourier-space fields such as $\tdelta(\vk,\tau)$.
This corresponds to Eq.(2.85) in \cite{Boyanovsky2008}, with a different time variable.
Equation (\ref{delta-Cs}) becomes identical to the standard equation for the linear density
modes in CDM scenarios when we set $c_s=0$ and $S=0$.
  
The new term in the left-hand side is similar to a pressure term
(at the linear level), and $c_s$ would be the comoving sound speed. The comoving Jeans
wave number $k_J$ would thus correspond to the comoving free-streaming wave number
$k_{\rm fs}$, with
\beq
k_{\rm fs}^2 = \frac{3\Om\cH^2}{2 c_s^2} .
\label{k-fs}
\eeq
However, this is only a formal analogy because there is no true pressure
as we consider a collisionless fluid. This term arises from the nonzero velocity dispersion
and its evolution with time is set by the comoving free-streaming wave number (\ref{k-fs})
rather than by a thermodynamical equation of state. 
At this linear level, $k_{\rm fs}(\tau)$ is obtained from the analysis of the linearized Vlasov 
equation. This yields \cite{Boyanovsky2008,Boyanovsky2011}
\beq
k_{\rm fs}(\tau) = k_{\rm fs}(\tau_{\rm eq})  \sqrt{ \frac{a}{a_{\rm eq}} } ,
\label{k-fs-eta}
\eeq
where $a_{\rm eq}$ is the scale factor at the matter-radiation equality, 
$a_{\rm eq}=1/(1+z_{\rm eq})$, with $z_{\rm eq} \simeq 3050$, and
\beq
k_{\rm fs}(\tau_{\rm eq}) \simeq \frac{11.17}{\sqrt{\bar{y}^2}} \left( \frac{m}{1 \rm keV}\right)
\left( \frac{g_d}{2} \right)^{1/2} \; {\rm Mpc}^{-1} ,
\label{k-fs-def}
\eeq
where $\bar{y}^2 \sim 10$ depends on the shape of the initial velocity distribution,
$g_d$ is the effective number of relativistic degrees of freedom at decoupling, and $m$ is
the mass of the dark matter particles.
 
The new source term in the right-hand side of Eq.(\ref{delta-Cs}) can be decomposed as
a memory term $S_{\rm NB}$ (i.e., an integral over past times) and an inhomogeneous
term $S_{\rm B}$ (i.e., an integral over the ``initial'' condition at matter-radiation equality);
see \cite{Boyanovsky2008,Boyanovsky2011}.
The nonlocal term $S_{\rm NB}$ decreases as $(k/k_{\rm fs})^4$ at low $k$, 
hence, it is subdominant on large scales as compared with the new term in the left-hand side.
The inhomogeneous term only decreases as $k^2$, but its relative importance decays with
time, as compared with the growing mode associated with the left-hand side.

In this paper, we neglect the source term $S$, and we investigate the equations of motion
\beqa
\frac{\pl \delta}{\pl\tau} + \nabla\cdot [(1+\delta) \vv] & = & 0 ,
\label{continuity} \\
\frac{\pl\vv}{\pl\tau} + \cH \vv + (\vv\cdot\nabla)\vv & = & - \nabla \Phi 
- \frac{c_s^2 \nabla\rho}{\rhob}  ,
\label{Euler}
\eeqa
where the gravitational potential is given by the Poisson equation (\ref{Poisson}), which
also reads as
\beq
\nabla^2 \Phi = \frac{3}{2} \Om \cH^2 \delta .
\label{Poisson-1}
\eeq
Equation (\ref{continuity}) is the usual continuity equation, which is exact when $\vv$ is
the mean peculiar velocity as defined by Eq.(\ref{v-def}). Equation (\ref{Euler}) is the
Euler equation, which only differs from the CDM case by the last term.
This equation is not exact, because as explained above this last term should be written in 
terms of the nonzero velocity dispersion, which obeys a third evolution equation that involves
the third-order velocity moment, and so on. The expression (\ref{Euler}) is the simplest
closure of this hierarchy that agrees with the linear theory (\ref{delta-Cs}) (where we neglect
the source term $S$). 
In a fluid analogy, where $c_s$ would be a uniform isothermal sound speed, this term
would be modified beyond linear order because it would read as $(\nabla \rho)/\rho$  
instead of $(\nabla\rho)/\rhob$.
However, because this analogy is not exact (and does not need to be strictly followed)
and to simplify the analysis, we keep the linear term of Eq.(\ref{Euler}).
This also ensures that we always mimic the slowdown of gravitational collapse on small
scales as compared with CDM. 
(In contrast, a truncation at a finite
order over $\delta$ of $\nabla \delta/(1+\delta)$ is not very well behaved. 
For instance, depending on whether the truncation order $p$ is odd or even
a term $(-1)^{p-1} \delta^{p-1} \nabla \delta$ either slows down or speeds up the dynamics of
overdense regions).

\subsection{Setup}
\label{Set-up}

The closure (\ref{Euler}) should be sufficient for our purpose, which is not to obtain the most
accurate predictions for large-scale statistics but to estimate the impact of a late-time 
velocity dispersion.
Equations (\ref{continuity})-(\ref{Euler}) read in Fourier space as
\beqa
\frac{\pl\tdelta}{\pl\tau}(\vk,\tau) + \ttheta(\vk,\tau) & = &
- \int\dd\vk_1\dd\vk_2 \; \delta_D(\vk_1+\vk_2-\vk) \nonumber \\
&& \times \alpha(\vk_1,\vk_2) \ttheta(\vk_1,\tau) \tdelta(\vk_2,\tau) ,
\label{F-continuity-1}
\eeqa
\beqa
\frac{\pl\ttheta}{\pl\tau}(\vk,\tau) + \cH \ttheta(\vk,\tau) + \frac{3\Om}{2}
\cH^2 [1+\epsilon(k,\tau)]  \tdelta(\vk,\tau) = \nonumber \\
&& \hspace{-8cm} - \!\! \int\dd\vk_1\dd\vk_2 \; \delta_D(\vk_1+\vk_2-\vk)
\beta(\vk_1,\vk_2) \ttheta(\vk_1,\tau) \ttheta(\vk_2,\tau) ,
\nonumber\\
&&
\label{F-Euler-1}
\eeqa
where we introduced the velocity divergence, $\theta=\nabla\cdot\vv$.
The usual kernels $\alpha$ and $\beta$ are given by
\beq
\alpha(\vk_1,\vk_2)= \frac{(\vk_1\!+\!\vk_2)\cdot\vk_1}{k_1^2} ,
\beta(\vk_1,\vk_2)= \frac{|\vk_1\!+\!\vk_2|^2(\vk_1\!\cdot\!\vk_2)}{2k_1^2k_2^2} ,
\label{alpha-beta-def}
\eeq
and the factor $\epsilon(k,\tau)$ writes as
\beq
\epsilon(k,\tau) = - \frac{k^2}{k_{\rm fs}(\tau)^2} .
\label{eps-def}
\eeq
Combining Eqs.(\ref{F-continuity-1}) and (\ref{F-Euler-1}) at linear order, we recover the
left-hand side of Eq.(\ref{delta-Cs}).
The equations of motion (\ref{F-continuity-1})-(\ref{F-Euler-1}) have the same form as
those studied in \cite{Brax2012} in the context of modified gravity models, but with a different
kernel $\epsilon(k,\tau)$. Therefore, we can use the same methods as in \cite{Brax2012} to
investigate large-scale structure formation.

To fully define our system, we must specify our initial conditions.
Following \cite{Viel2005}, we write the linear density power spectrum today in terms
of the reference CDM power spectrum as
\beq
P_{L,\rm WDM}(k,z=0) = P_{L,\rm CDM}(k,z=0) \; T(k)^2 ,
\label{P-WDM}
\eeq
with
\beq
T(k) = [1+(\alpha k)^{2\nu}]^{-5/\nu} ,
\label{T-def}
\eeq
with $\nu=1.12$ and
\beq
\alpha = 0.049 \left( \frac{m}{1 \rm keV} \right)^{-1.11} \left( \frac{\Om}{0.25} \right)^{0.11}
\left( \frac{h}{0.7} \right)^{1.22} \; h^{-1} \; \mbox{Mpc} .
\label{alpha-T-def}
\eeq
The high-$k$ cutoff (\ref{T-def}) is due to free streaming in the relativistic regime,
which damps the growth of small-scale clustering as compared with CDM.

Then, in a manner similar to what is done in numerical simulations, we choose an
``initial'' redshift $z_i = 100$, and we write the initial power spectrum as
$P_L(k,z_i) = [D_+(z_i)/D_+(0)] P_{\rm WDM}(k,0)$, where $D_+(z)$ is the CDM
linear growing mode, which is independent of $k$ (it is the growing solution
of Eq.(\ref{delta-Cs}) with $c_s=0$ and $S=0$).
Next, for a given initial redshift $z_i$, we compute the subsequent formation of large-scale
structure defined by the equations of motion (\ref{F-continuity-1})-(\ref{F-Euler-1})
with either $\epsilon\neq 0$ (i.e., taking into account the late-time velocity dispersion
through the last term in Eq.(\ref{Euler})) or $\epsilon= 0$ (i.e., neglecting this late-time
velocity dispersion), starting with the same initial WDM power spectrum (\ref{P-WDM}).
Comparing these two results we obtain an estimate of the impact of the nonzero
velocity dispersion at low redshifts, below $z_i$.
Comparing with a pure CDM scenario, with the initial power spectrum $P_{\rm CDM}$,
we compare this effect with the damping due to the initial high-$k$ cutoff (\ref{T-def}),
which mostly arises from the relativistic era.
This will allow us to check whether the late-time dark matter velocity dispersion can be
neglected on large scales (outside of collapsed objects where the upper bound associated
with Liouville theorem may be important). This is often the assumption used in numerical
simulations, where the system is modeled with the same codes as for CDM and
the only difference lies in the initial condition at redshift $z_i$, that is, in the damping
(\ref{T-def}) of the initial power spectrum 
\cite{Colombi1996a,Bode2001,Schneider2011,Dunstan2011}.

In some cases, one adds to the macroparticles used in the N-body simulations
an additional initial random velocity, $v_{\rm rms}$, drawn from the thermal distribution
$f(v)$ of the WDM particles \cite{Colin2008,Boyarsky2009b,Viel2012}.
However, this is not fully legitimate because
these macroparticles have a much larger mass ($\sim 10^5 M_{\odot}$) than the
WDM particles and a clump of this mass of many elementary dark matter particles would
have a much smaller (almost zero) mean velocity.
In fact, setting such initial conditions at $z_i=100$ leads to spurious power at high $k$ in
the power spectrum measured at later times (starting at $z_i=40$ appears to avoid this
problem because $v_{\rm rms}$ is smaller) \cite{Colin2008}.
This can be understood from the fact that adding these random velocities is equivalent
to model a CDM scenario, with the damped power spectrum (\ref{P-WDM}), to which is
added a $k^2$ high-$k$ tail associated with the random velocity component.
Indeed, this adds a white-noise component to the initial velocity power spectrum,
which corresponds to a $k^2$ tail for the density power spectrum (when we decompose
over growing and decaying modes).
On the other hand, these additional random velocities can be seen as an effective
tool to build an upper bound on the phase-space distribution function and to investigate
the formation of cored profiles associated with Liouville theorem \cite{Maccio2012}.
However, this shows that it is not easy to include the WDM velocity dispersion in
numerical simulations in a realistic fashion.
This is another motivation for the analytic study presented in this paper.
 
In the following, as in \cite{Schneider2011}, for numerical computations we adopt a 
background cosmology that is consistent with WMAP7 \cite{Komatsu2011},
$\Omega_{\rm m}=0.2726$, $\Omega_{\Lambda}=0.7274$, $\Omega_{\rm b}=0.046$,
$h=0.704$, $n_s=0.963$, and $\sigma_8=0.809$.

We focus on the case of a 1~keV dark matter particle, and we take $\bar{y}^2=12.939$
as for thermal fermions (or sterile neutrinos produced via the Dodelson-Widrow 
\cite{Dodelson1994} nonresonant mixing mechanism) and $g_d=10.75$
\cite{Boyanovsky2011,Viel2005}.
This gives $k_{\rm fs}(\tau_{\rm eq}) \simeq 10 h$Mpc$^{-1}$. 
The comoving free-streaming distance, traveled by particles since $t_{\rm eq}$ because
of thermal velocities,
\beq
\lambda_{\rm fs}(t) = \int_{t_{\rm eq}}^t \frac{\dd t'}{a(t')} c_s(t') ,
\label{lambda-def}
\eeq
converges at late times to \cite{Boyanovsky2008}
\beq
t \gg t_{\rm eq} : \;\; \lambda_{\rm fs} \simeq \frac{\sqrt{6}}{k_{\rm fs}(\tau_{eq})} \simeq
0.24 h^{-1} \mbox{Mpc}
\eeq
This gives the scale below which the velocity dispersion of the warm dark matter
has a strong effect.

\section{Results}
\label{Results}

\subsection{Matter density power spectrum}
\label{power-spectrum}

\begin{figure}
\begin{center}
\epsfxsize=8.5 cm \epsfysize=6 cm {\epsfbox{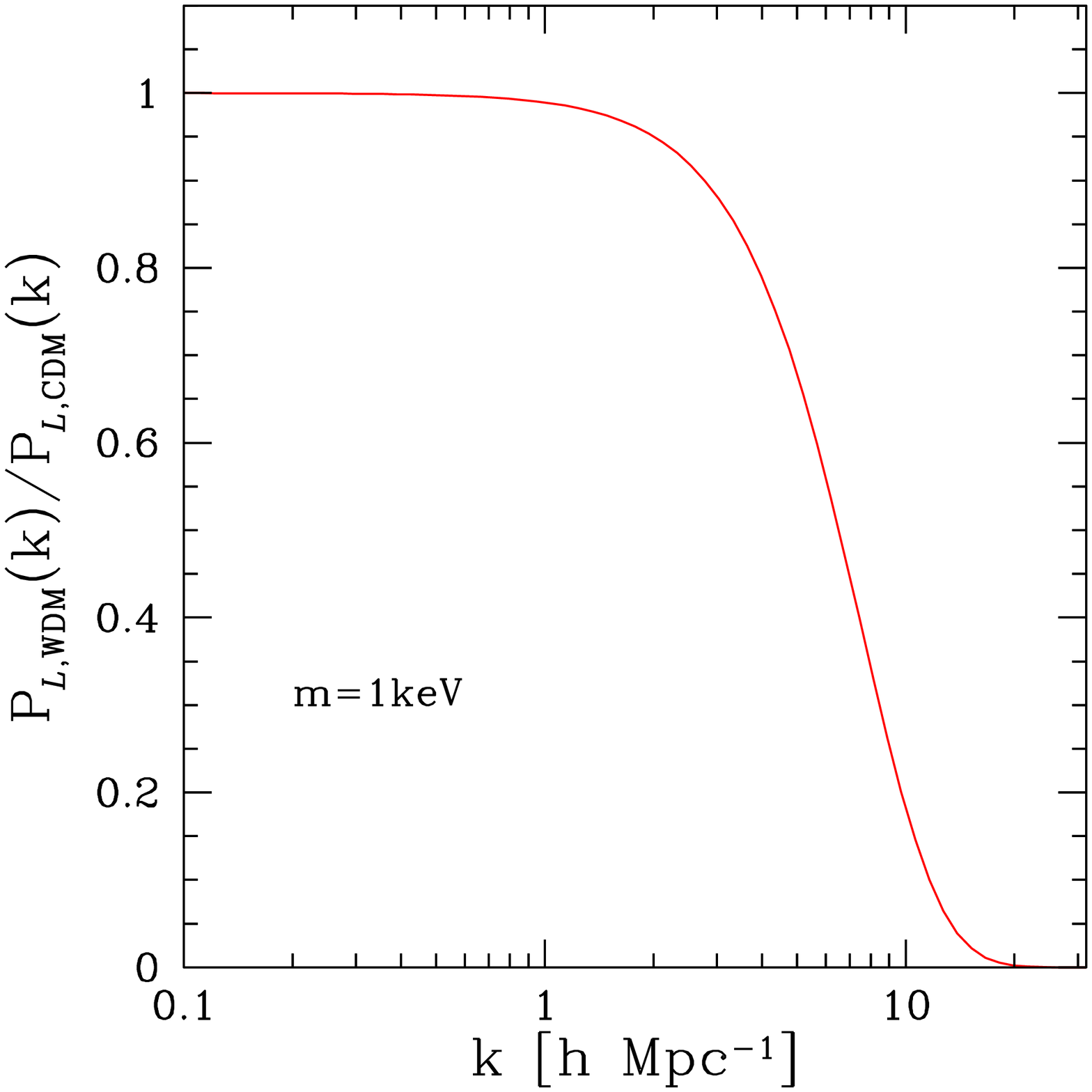}}
\epsfxsize=8.5 cm \epsfysize=6 cm {\epsfbox{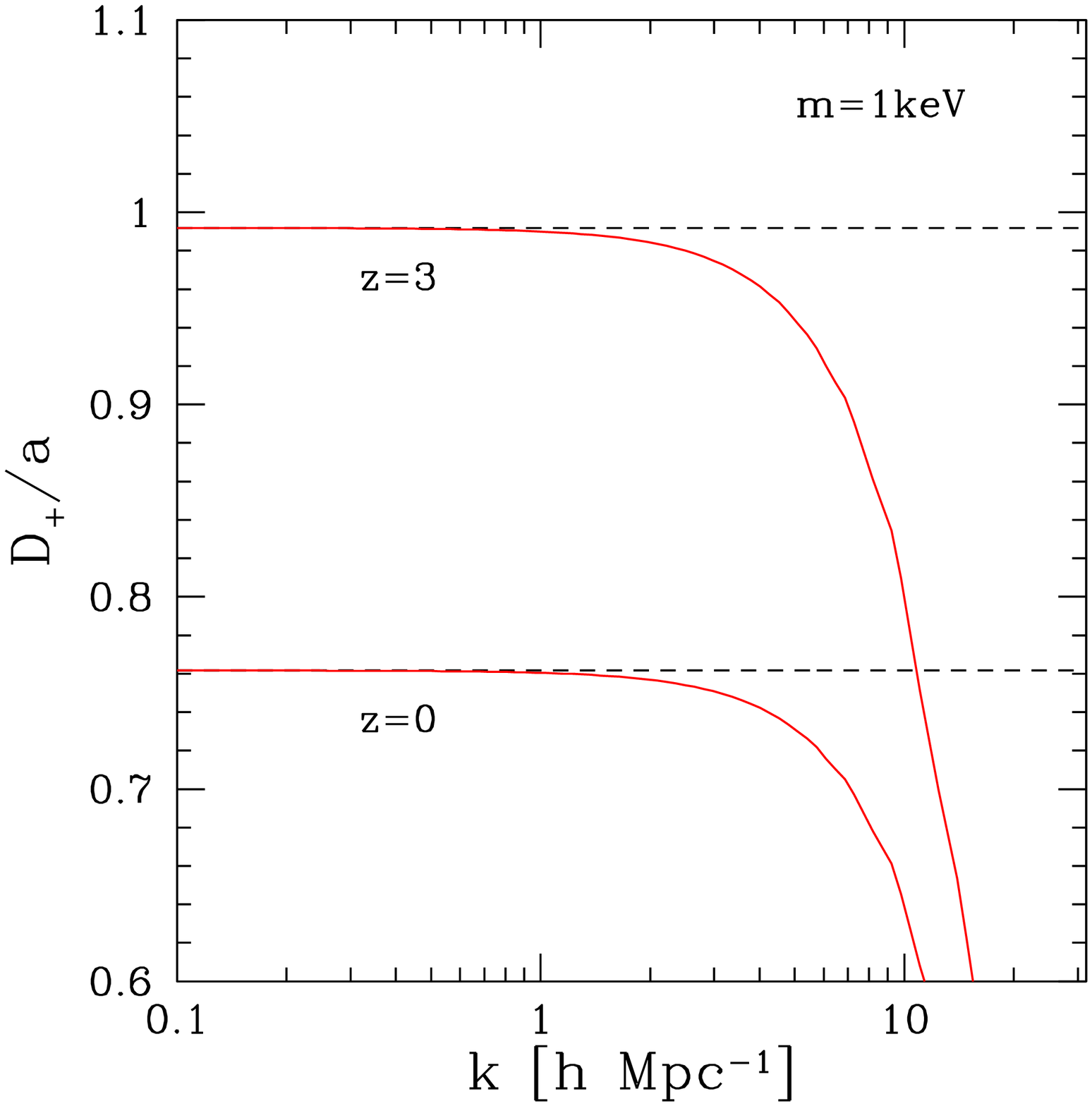}}
\end{center}
\caption{{\it Upper panel:} ratio $T(k)^2= P_{L,\rm WDM}/P_{L,\rm CDM}$
of the linear WDM and CDM power spectra at $z=0$, from Eq.(\ref{P-WDM}).
{\it Lower panel:} linear growing mode $D_+(k,\tau)$ normalized to the scale factor
$a(\tau)$ as a function of the wave number. We show our results at $z=0$ and $z=3$
for the CDM case (black dashed lines) and the WDM case (red solid lines) with $m=1$
keV.}
\label{fig-Dplin_k}
\end{figure}

Since the equations of motion (\ref{F-continuity-1})-(\ref{F-Euler-1}) have the same
form as those studied in \cite{Brax2012} in the context of modified gravity models (but with a different
kernel $\epsilon(k,\tau)$) we use the same methods as in \cite{Brax2012} for our numerical
computations. We refer the reader to \cite{Brax2012} for a description of our analytical methods.

Because of the explicit dependence on $k$ introduced in the left-hand side of the
Euler equation (\ref{F-Euler-1}) by the factor $\epsilon(k,\tau)$, the linear
growing and decaying modes of the density contrast now depend on $k$.
As explained above, they also satisfy Eq.(\ref{delta-Cs}), with $S=0$, where the
dependence on $k$ is explicit. This pressurelike term, $-k^2 c_s^2$, slows down the
growth of density perturbations at high $k$. For $k > k_{\rm fs}$, which plays the
role of a Jeans wave number, density perturbations would no longer grow but oscillate.
We show in the lower panel of Fig.~\ref{fig-Dplin_k} the linear growing mode
$D_+(k,\tau)$ as a function of the wave number. 
(We normalize all linear growing modes to the CDM mode at the initial redshift $z_i$.)
We clearly see the decrease of the growing
mode above $k \sim 1/\lambda_{\rm fs} \sim 4 h$ Mpc$^{-1}$. 
For comparison, we also plot in the upper panel
the ratio $P_{L,\rm WDM}/P_{L,\rm CDM}=T(k)^2$ of the linear WDM to CDM power
spectra, from Eq.(\ref{P-WDM}). All curves deviate from the CDM prediction at about the
same wave number, but, as expected, the damping of the linear power spectrum is stronger
than the damping of the late-time linear growing mode.
Indeed, the damping of the linear power spectrum shown in the upper panel is mostly
due to early-time effects, when the dark matter particles were still relativistic.
In contrast, by definition of our initial conditions at $z_i=100$, the damping found in the
lower panel is a late-time effect at $z<z_i$, due to the small nonzero velocity dispersion.
This effect declines with time as the comoving wave number $k_{\rm fs}(\tau)$
grows as $\sqrt{a}$ from Eq.(\ref{k-fs-eta}). However, it is not zero and we will
estimate in the following the magnitude of this late-time effect.

\begin{figure*}
\begin{center}
\epsfxsize=6.9 cm \epsfysize=5.4 cm {\epsfbox{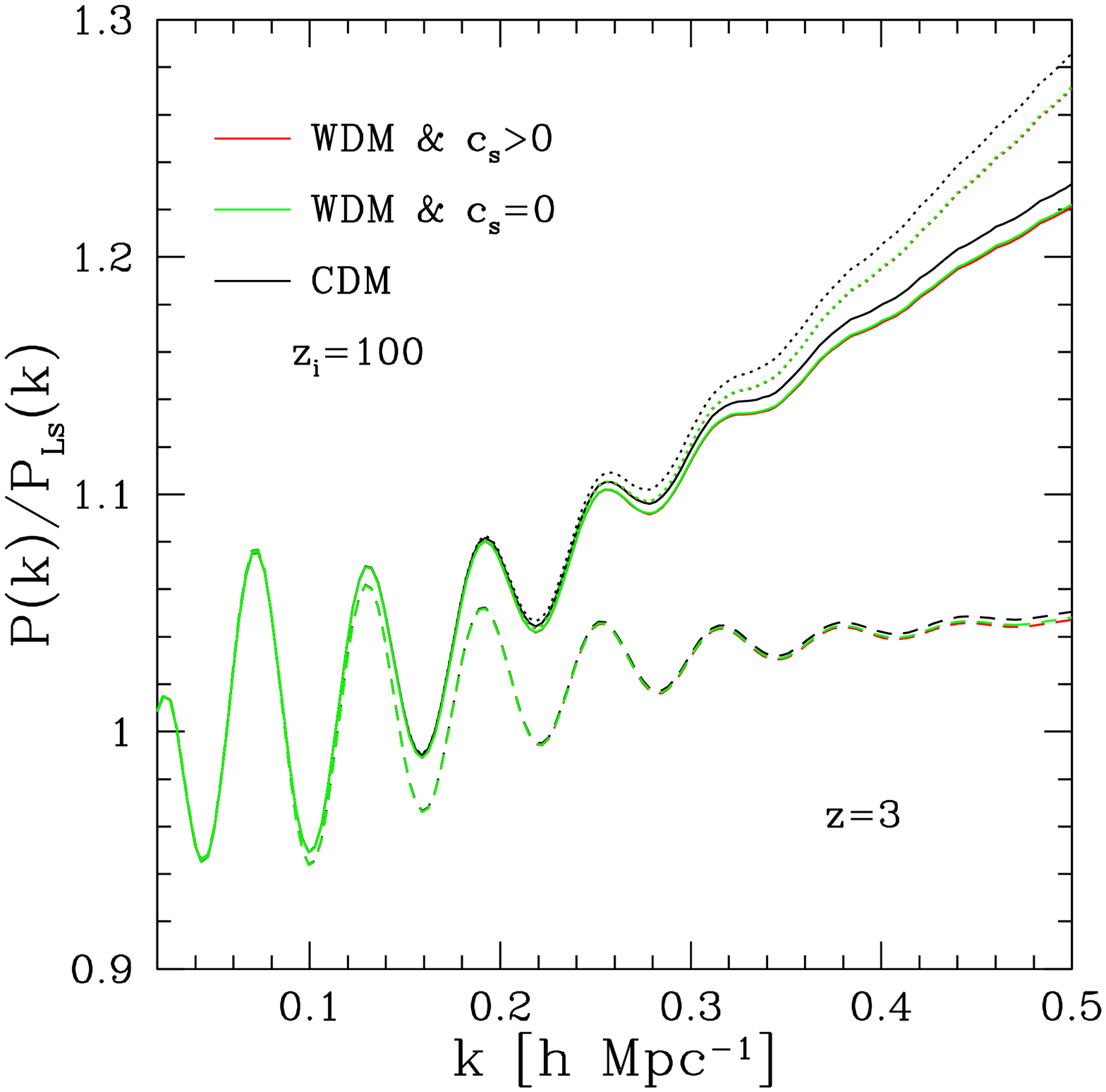}}
\epsfxsize=6.9 cm \epsfysize=5.4 cm {\epsfbox{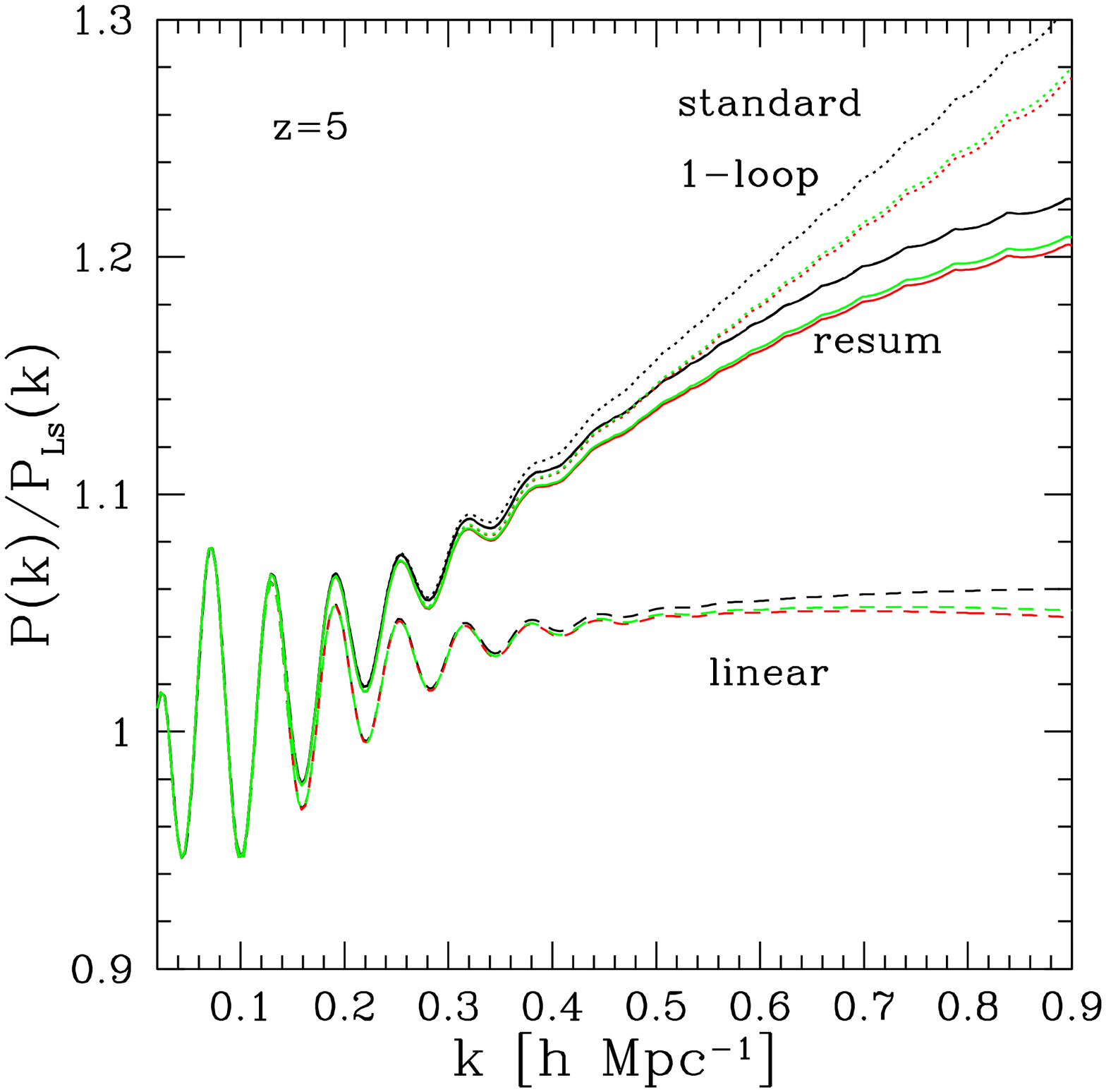}}
\end{center}
\caption{Ratio of the power spectrum $P(k)$ to a smooth
$\Lambda$CDM linear power spectrum $P_{Ls}(k)$ without baryonic oscillations,
from \cite{Eisenstein1999}. We plot the linear power (lower group of dashed lines), the
nonlinear one-loop ``steepest descent'' resummation (middle group of solid lines), and the
``standard'' 1-loop result (upper group of dotted lines).
In each case, we show our results for the reference CDM scenario
(slightly upper black lines), the WDM scenario with $c_s=0$ (middle green lines) and
with $c_s\neq 0$ (slightly lower red lines).}
\label{fig-BAO}
\end{figure*}

\begin{figure*}
\begin{center}
\epsfxsize=7.2 cm \epsfysize=5.4 cm {\epsfbox{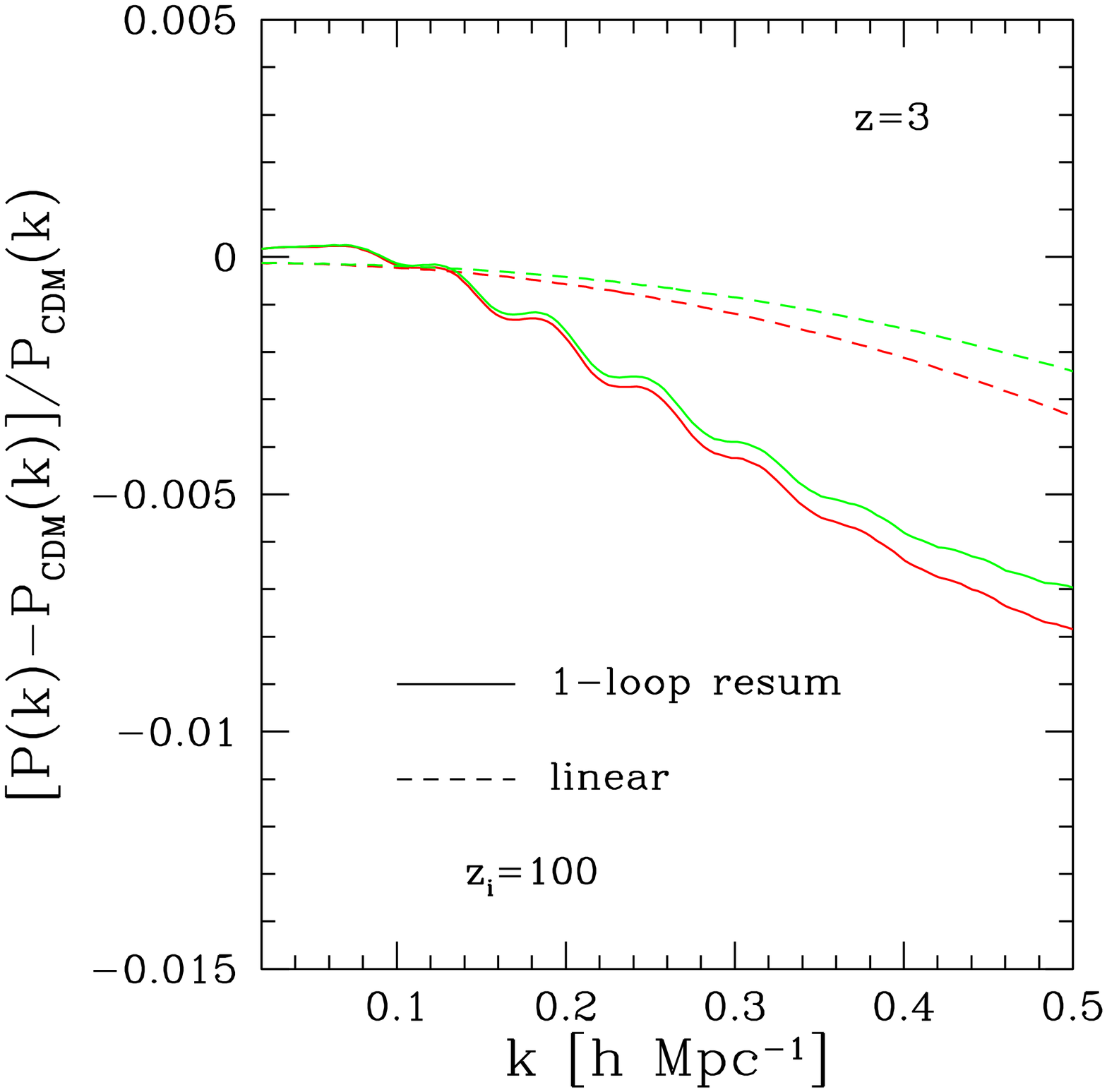}}
\epsfxsize=7.2 cm \epsfysize=5.4 cm {\epsfbox{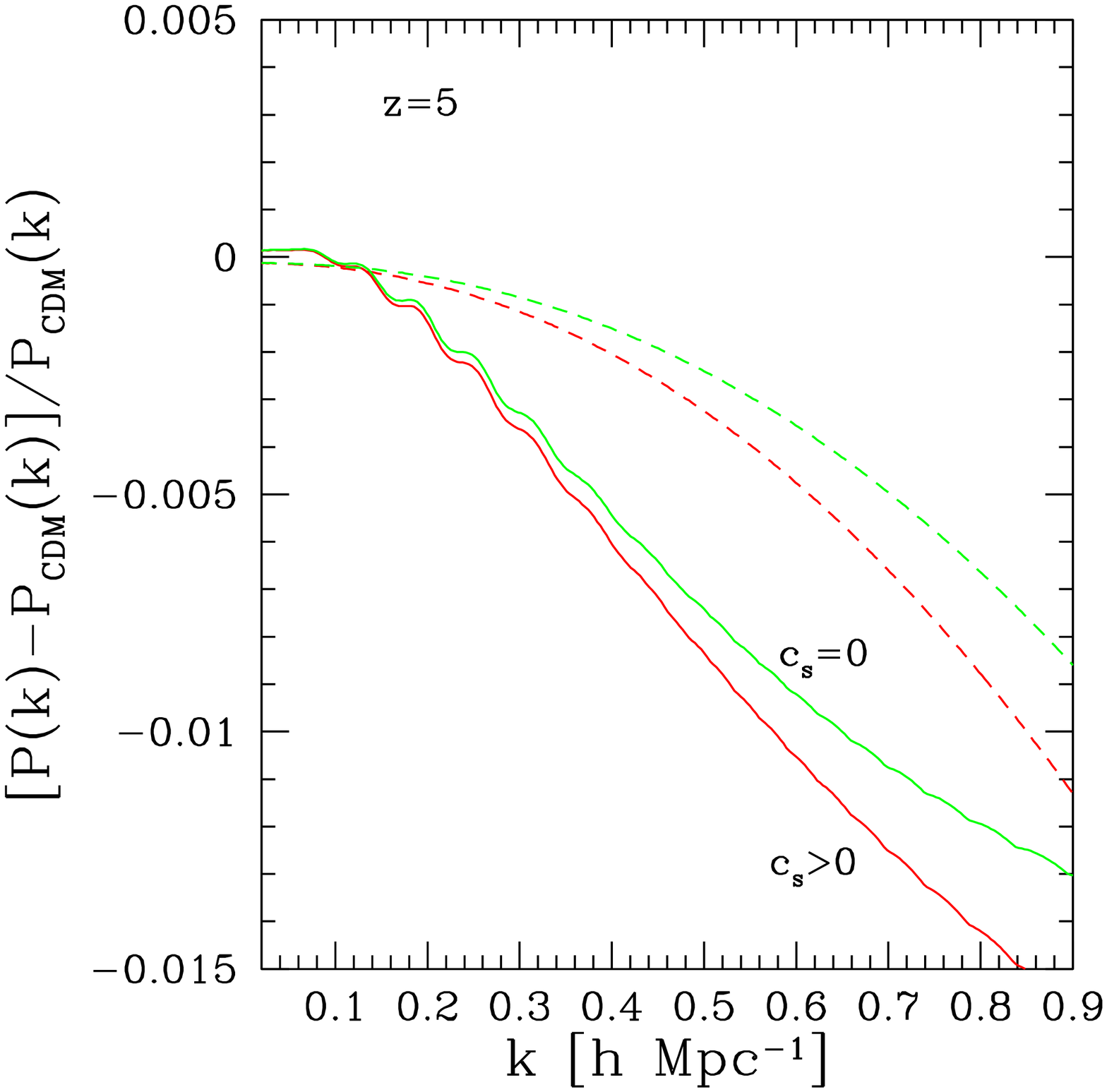}}
\end{center}
\caption{Relative deviation of the power spectrum from the CDM reference, in terms of the
linear power spectra ($\Delta P_L/P_L$, upper group of dashed lines), and of the
nonlinear power spectra ($\Delta P/P$, lower group of solid lines) obtained
from the one-loop ``steepest descent'' resummation.
In each case, we show our results for the WDM scenarios with $c_s=0$ (slightly upper
green lines) and with $c_s\neq 0$ (slightly lower red lines).}
\label{fig-dPk}
\end{figure*}

Next, we consider the nonlinear density power spectrum in the perturbative regime
in Figs.~\ref{fig-BAO} and \ref{fig-dPk}.
We compare the results associated with the reference $\Lambda$CDM scenario,
the WDM scenario with $c_s=0$ (i.e., the only difference from CDM arises from the initial
power (\ref{P-WDM})), and the WDM scenario with $c_s \neq 0$ (i.e., the difference
from CDM arises from both the initial power (\ref{P-WDM}) and the late-time
velocity-dispersion).
To emphasize the difference between various curves, we plot the ratios of our results
by a common reference linear power spectrum without baryonic oscillations,
from \cite{Eisenstein1999}, in Fig.~\ref{fig-BAO}
We plot both the ``standard'' one-loop perturbative result \cite{Bernardeau2002}
and the ``steepest-descent resummation'' \cite{Valageas2007,Valageas2008,Valageas2011d}, 
which agrees with the standard result up to one loop and contains
a partial resummation of higher-order terms 
\footnote{To avoid spending most of the computation time on high $k$ wavenumbers where
the initial power is very small, as seen in Fig.~\ref{fig-Dplin_k}, we set a lower bound
$\epsilon(k,\tau) \geq -0.5$ in the numerical computations.}.

We recover the suppression of the nonlinear power spectrum due to the
high-$k$ cutoff (\ref{T-def}) \cite{Dunstan2011}.
As in the usual CDM case, we can see that the nonlinear dynamics amplifies the
power spectrum while erasing some of the baryonic oscillations.
The difference between the ``standard'' and the ``resummed'' perturbative predictions
is similar to the one obtained in the CDM scenario, and it is larger than the difference
between the CDM and WDM results. This means that on these scales, the standard
perturbation theory is not accurate enough to describe the deviations between the
CDM and WDM predictions (for $m \geq 1$keV), which are on the order of $1\%$
as seen in Fig.~\ref{fig-dPk}.

\begin{figure*}
\begin{center}
\epsfxsize=5.9 cm \epsfysize=5.4 cm {\epsfbox{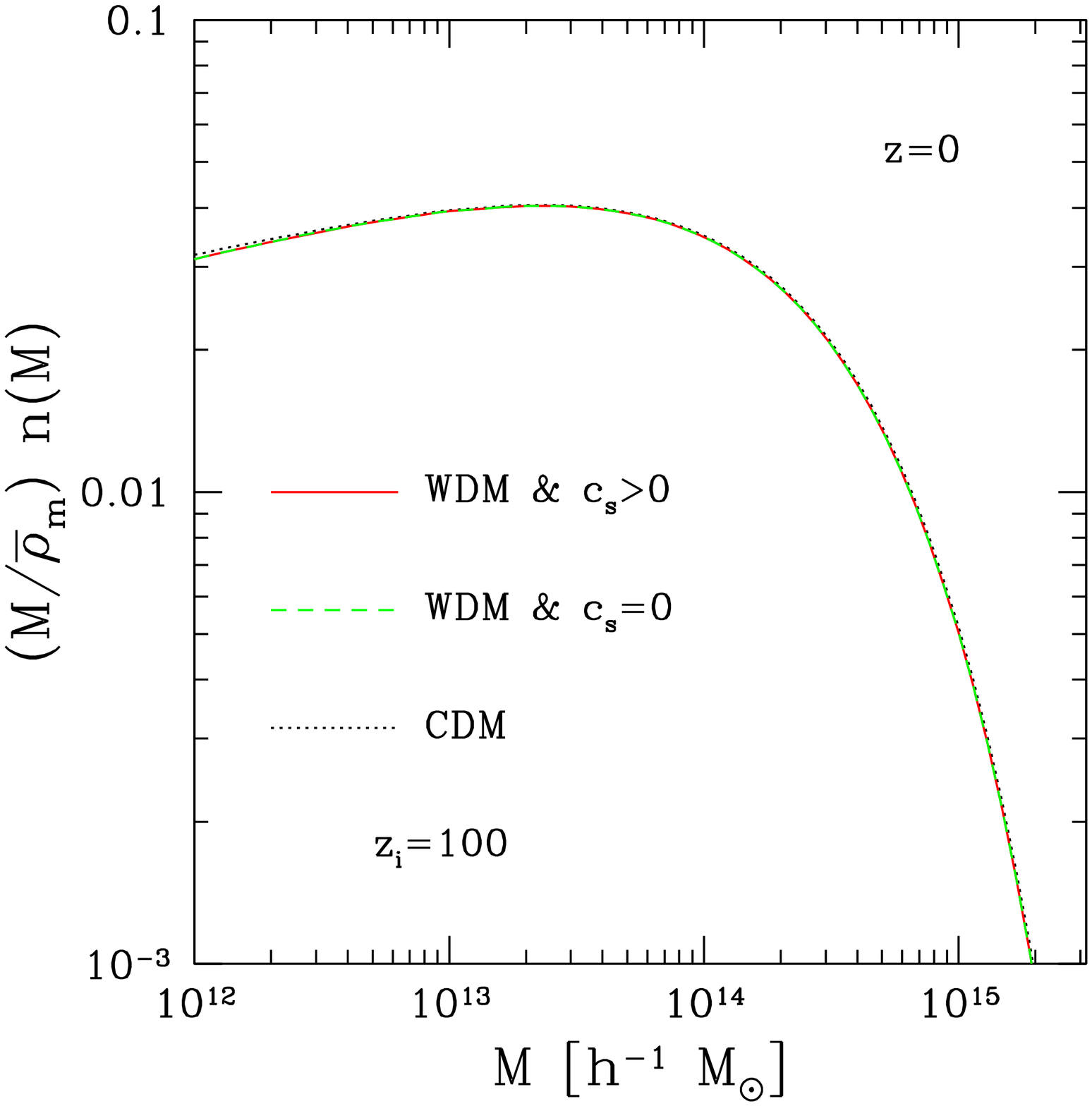}}
\epsfxsize=5.9 cm \epsfysize=5.4 cm {\epsfbox{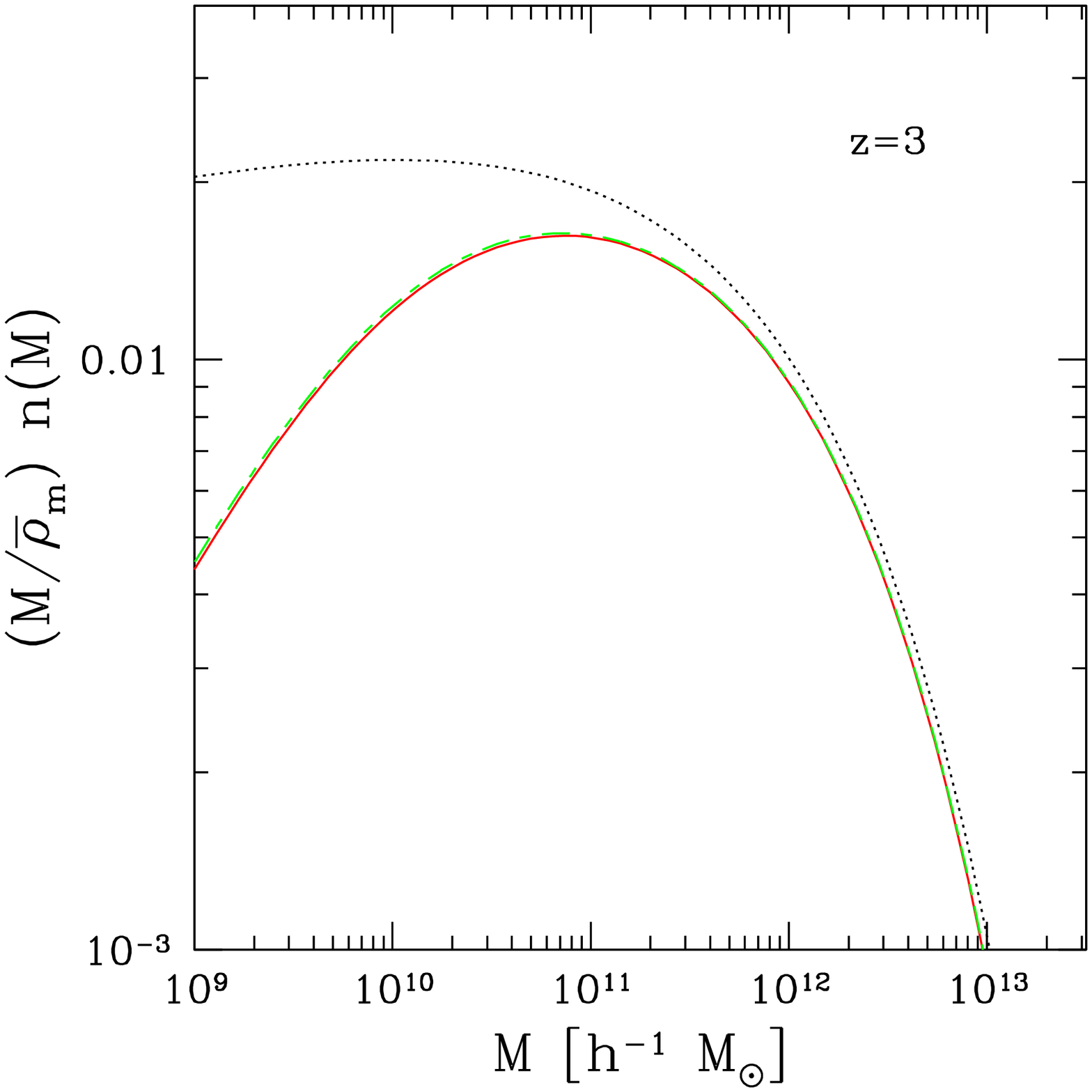}}
\epsfxsize=5.9 cm \epsfysize=5.4 cm {\epsfbox{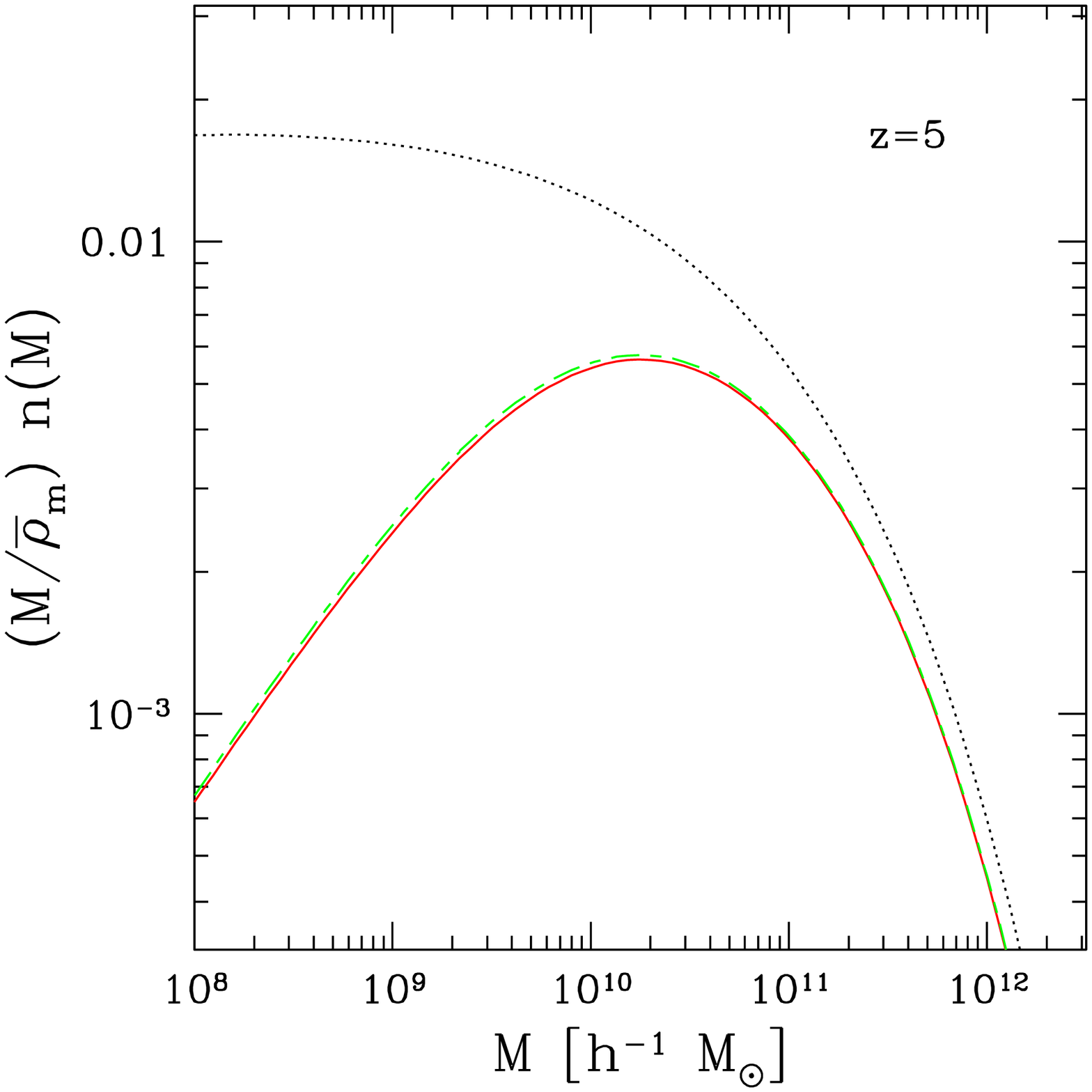}}
\end{center}
\caption{Halo mass functions for the reference CDM scenario
(black dotted lines), the WDM scenarios with $c_s=0$ (green dashed lines), and with
$c_s\neq 0$ (red solid lines).}
\label{fig-lnM}
\end{figure*}

On the other hand, we can see that nonlinear contributions amplify these deviations, as
compared with the linear power spectra.
This is more clearly seen in Fig.~\ref{fig-dPk}, where we plot the relative deviations from
the reference linear and nonlinear CDM power spectra.
The comparison of the curves obtained with $c_s=0$ and $c_s \neq 0$ shows that
most of the damping with respect to the CDM case is due to the cutoff (\ref{T-def})
of the initial power spectrum. The late-time velocity dispersion only slightly amplifies
the damping on these scales.
Thus, for practical purposes, this late-time effect may be neglected. This justifies the
use of N-body simulations, initially built for CDM, to study gravitational clustering
on these scales \cite{Schneider2011,Dunstan2011}.
However, for $m\geq 1$keV, the difference from the CDM power is not larger than
$1.5\%$ on these scales. This means that, as expected, most of the constraints that
can be set from observations on the WDM scenario arise from smaller scales.
However, because these smaller scales are also more difficult to predict
(since they are deeper in the nonlinear or nonperturbative regime), it is interesting
to check the signal that can be expected on the larger perturbative scales shown
in Figs.~\ref{fig-BAO} and \ref{fig-dPk}, which are better controlled since they can be
described by systematic perturbative schemes.

\subsection{Halo mass function}
\label{mass-function}

Another key statistic of large-scale structures is the mass function of collapsed
halos. As in \cite{Brax2012} we define halos by a nonlinear density contrast of $200$. Then,
we obtain the halo mass function from the spherical dynamics, using the 
Press-Schechter scaling variable $\nu$ \cite{Press1974},
\beq
n(M) \frac{\dd M}{M} = \frac{\rhob_{\rm m}}{M} \; f(\nu) \; \frac{\dd\nu}{\nu} ,
\label{n-M}
\eeq
with
\beq
\nu = \frac{\cF_q^{-1}(200)}{\sigma_q} ,
\label{nu-def}
\eeq
and the scaling function $f(\nu)$ from \cite{Valageas2009}
\beq
f(\nu)= 0.502 \left[ (0.6 \nu)^{2.5} + (0.62 \nu)^{0.5} \right] \, e^{-\nu^2/2} ,
\label{f-nu}
\eeq
which has been fitted to $\Lambda$CDM numerical simulations.
This ensures that the halo mass function is always normalized to unity and
obeys the large-mass tail $n(M) \sim e^{-\nu^2/2}$ for any spherical-collapse
mapping $\cF_q$.
We obtain the spherical dynamics associated with the equations of motion
(\ref{F-continuity-1})-(\ref{F-Euler-1}) as in \cite{Brax2012}.
This provides the spherical-collapse mapping,
$\delta_{Lq} \mapsto \delta_x =\cF_q(\delta_{Lq})$, from the linear density contrast
on the Lagrangian radius $q$ to the nonlinear density contrast on the Eulerian radius $x$.

We show our results in Fig.~\ref{fig-lnM}. The low-mass tail should be considered with
caution because its exponent may depend on the shape of the linear power spectrum
and be different from the CDM case. 
Thus, numerical simulations suggest that a simple recipe of the form (\ref{n-M}), which
involves a scaling function $f(\nu)$ fitted to CDM simulations, overestimates the
low-mass tail in WDM scenarios \cite{Schneider2011}.
In contrast, the large-mass tail is better
controlled because it is governed by spherically symmetric saddle points
(and it does not involve the multiple mergers that affect the low-mass tail).

As is well known, the WDM scenario leads to a much smaller halo mass function
at low masses than in the CDM case, because of the lack of power on small scales
\cite{Bode2001,Schneider2011,Dunstan2011}.
For our purposes, Fig.~\ref{fig-lnM} shows that the impact of the late-time velocity
dispersion on the halo mass function at $z\leq 5$ can be neglected for $m\geq 1$keV.
Indeed, the difference between the two WDM curves associated with either $c_s\neq 0$
or $c_s=0$ at $z\leq z_i$ is much smaller than the deviation from the CDM reference.
It is also smaller than the accuracy of halo mass functions that can be obtained from
phenomenological models or numerical simulations.
Again, this means that standard N-body codes can be used to predict the halo mass
function for WDM scenarios (with $m\geq 1$keV).

This agrees with results from \cite{Barkana2001}, which also model the velocity dispersion
as an effective pressure term and compute the delay of halo collapse by this
hydrodynamical-like pressure for a spherical dynamics.
They also find that the initial cutoff of the linear power spectrum plays a greater role
than halo dynamics, but the latter becomes significant at very low mass and high
redshift (the late-time velocity dispersion has a very small effect below $z<40$).

\begin{figure*}
\begin{center}
\epsfxsize=5.9 cm \epsfysize=5.4 cm {\epsfbox{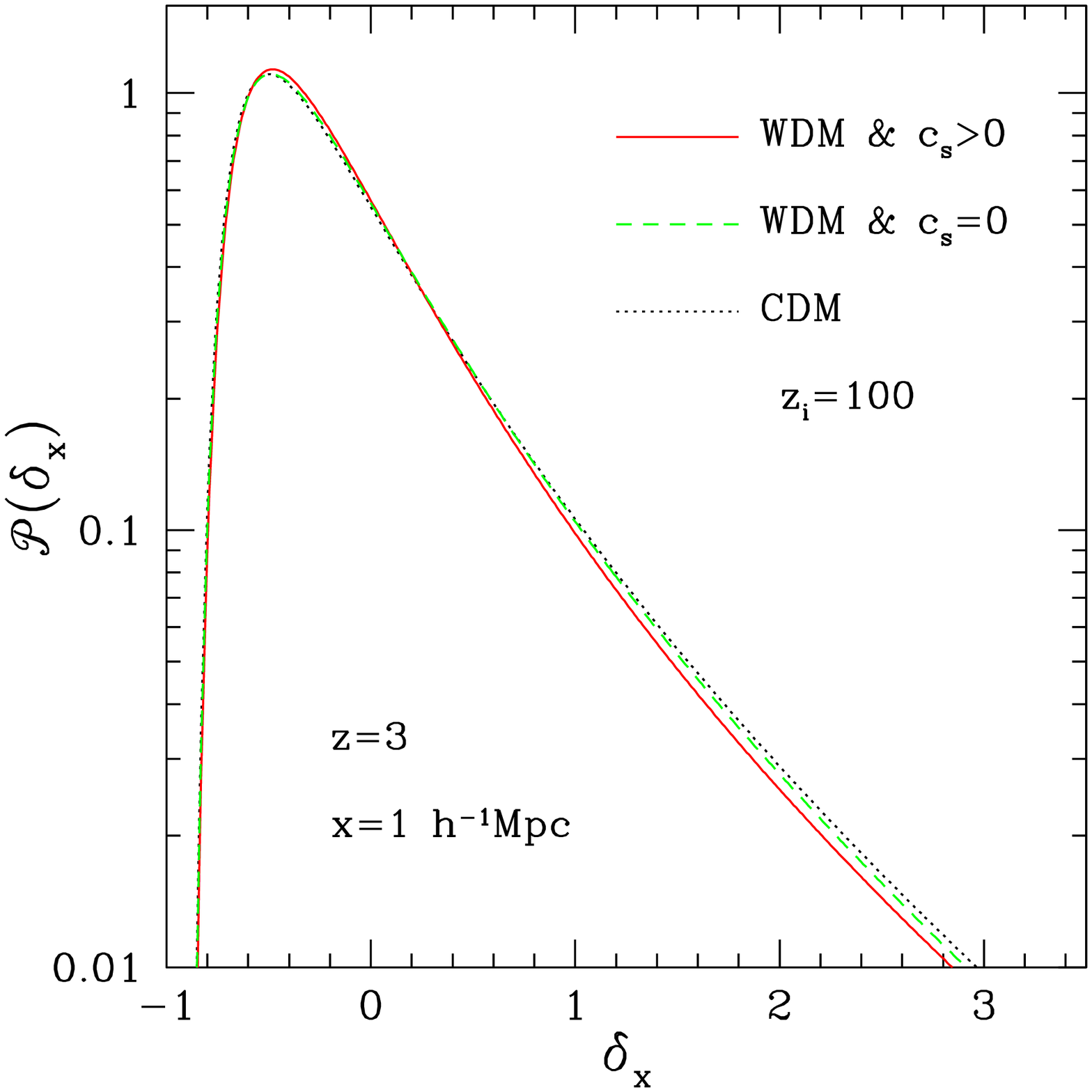}}
\epsfxsize=5.9 cm \epsfysize=5.4 cm {\epsfbox{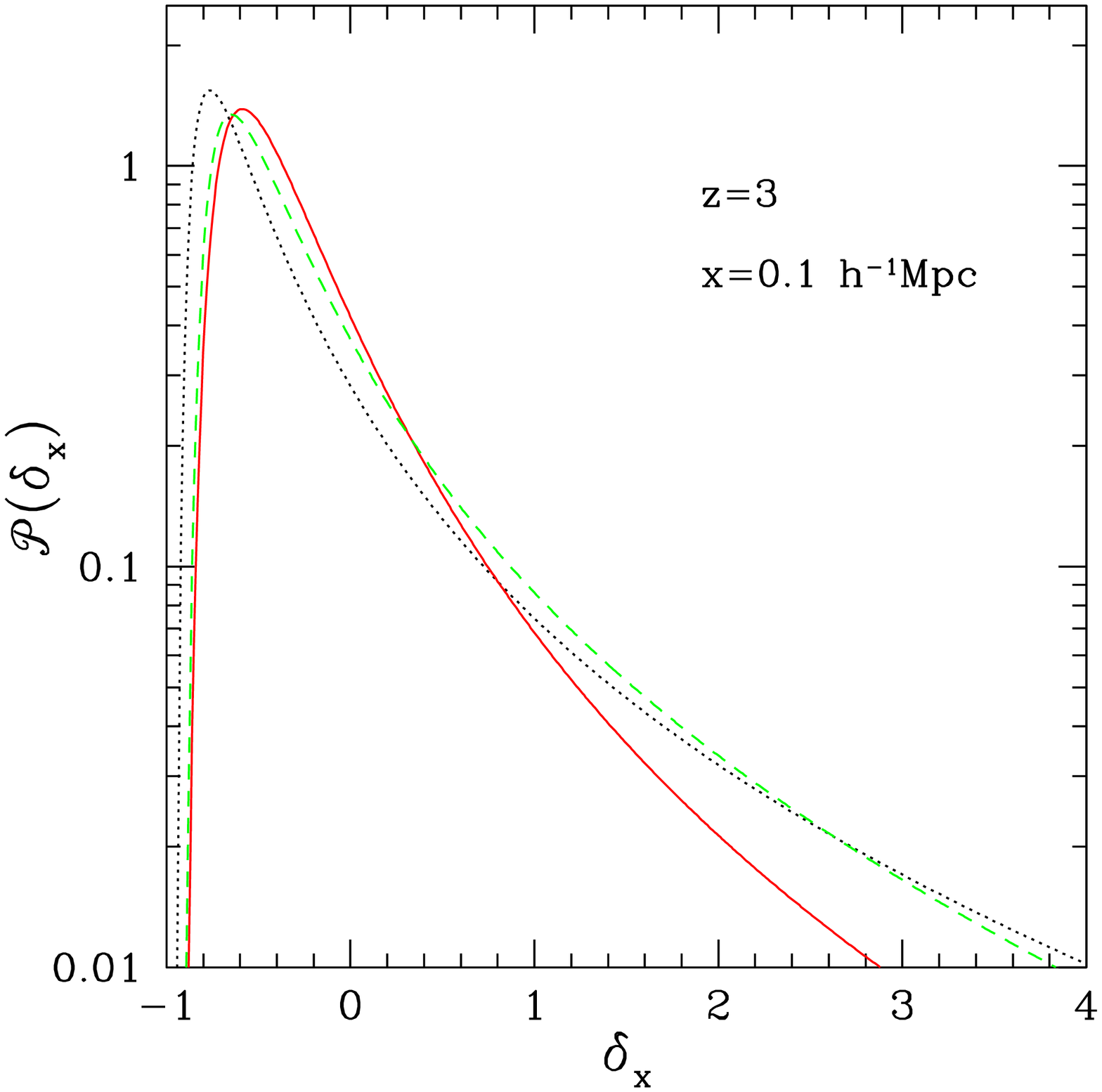}}
\epsfxsize=5.9 cm \epsfysize=5.4 cm {\epsfbox{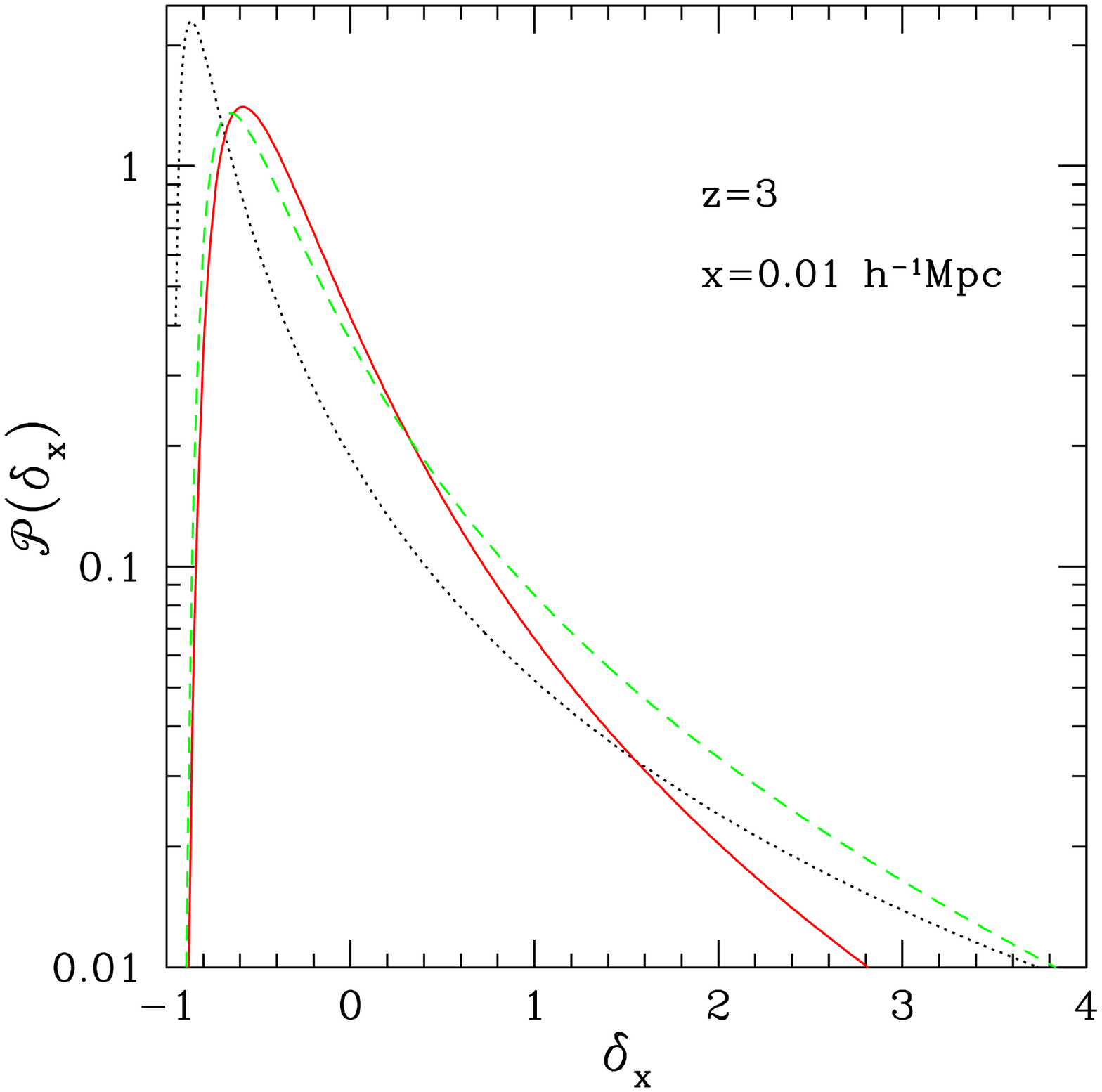}}
\end{center}
\caption{Probability distribution $\cP(\delta_x)$ of the matter density contrast
within spheres of radius $x$, for three radii. 
We show our results at $z=3$ for the reference CDM
scenario (black dotted lines), the WDM scenarios with $c_s=0$ (green dashed lines),
and with $c_s\neq 0$ (red solid lines).}
\label{fig-lPrho}
\end{figure*}

Figure~\ref{fig-lnM} can be compared with Fig.~7 of \cite{Benson2012}, which estimated the
WDM halo mass function in the excursion-set formalism.
These authors also find that the effect of the velocity dispersion, if it is estimated from its
impact on the critical overdensity, is very small on the relevant mass scales 
(so that the cutoff of the halo mass function at low mass is set by the initial cutoff of the
power spectrum (\ref{T-def})).
However, they find that if one modifies in addition the scaling function
$f(\nu)$, to take into account the dependence of the mass
function on the global shape of the function $\delta_c(M)$ (and not only on its value at the
mass $M$ of interest), the low-mass cutoff becomes sharper.
This second effect is not included in our paper (it requires additional modeling, such as the 
excursion set formalism, which needs to be checked
against numerical simulations in this regime).

\subsection{Probability distribution of the density contrast}
\label{Probability}

Finally, we consider the probability distribution $\cP(\delta_x)$ of the density contrast
within spheres of radius $x$.
As in \cite{Brax2012}, we use a steepest-descent approach to obtain $\cP(\delta_x)$ from the
spherical dynamics \cite{Valageas2002,Valageas2009}.
This is valid in the mildly nonlinear regime.
We plot our results in Fig.~\ref{fig-lPrho} at $z=3$, on scales that correspond to
Lyman-$\alpha$ clouds.
Indeed, depending on the details of the models Lyman-$\alpha$ clouds are associated
with scales from $\sim 10 h^{-1}$kpc to $\sim 1 h^{-1}$Mpc (from the small 
Lyman-$\alpha$ forest clouds to damped systems)
\cite{Valageas1999b,Bi1993,Rees1986}.

We recover the characteristic asymmetry induced by the nonlinear gravitational
dynamics, with a shift of the peak toward underdensities (most of the volume is
underdense), a very sharp low-density cutoff ($\delta_x\geq -1$ since the matter
density is always positive), and an extended high-density tail (most of the mass is
within overdensities).
On large scales, $x \geq 1 h^{-1}$Mpc, all curves are very close and the deviations
between the CDM and WDM scenarios are small.
On smaller scales, the lack of small-scale power in the WDM scenario leads to a
less advanced stage of the nonlinear evolution: the peak shifts closer to the mean
$\langle\delta_x\rangle=0$, and the tails are sharper.
The late-time velocity dispersion even further impedes the nonlinear evolution and
makes the large density tail sharper.
This suggests that accurate measures of the probability distribution of the flux decrement
of distant quasars, due to Lyman-$\alpha$ absorption lines, which is closely related to the 
probability distribution of the matter density on these scales \cite{Valageas1999b,Viel2002}, 
could be sensitive to this late-time velocity dispersion.
Thus, numerical simulations that do not include this effect are likely to underestimate
somewhat the difference between the CDM and WDM scenarios with respect to
Lyman-$\alpha$ absorption lines.

It is not surprising that these statistics are more sensitive to the late-time velocity dispersion than
the quantities studied in previous figures (the power spectrum on large perturbative scales
and the halo mass function). Indeed, it is well-known that, because they probe relatively
small scales, Lyman-$\alpha$ clouds are a sensitive probe of WDM scenarios.
Most works focus on the decrease of the flux power spectrum due to the high-$k$
damping of the WDM power spectrum set by the relativistic free streaming
\cite{Viel2005,Abazajian2006a,Seljak2006}.
Figure~\ref{fig-lPrho} shows that the late-time velocity dispersion also has a non-negligible
effect.

\subsection{Range of validity of our approximations}
\label{Range}

\begin{figure*}
\begin{center}
\epsfxsize=5.9 cm \epsfysize=5.4 cm {\epsfbox{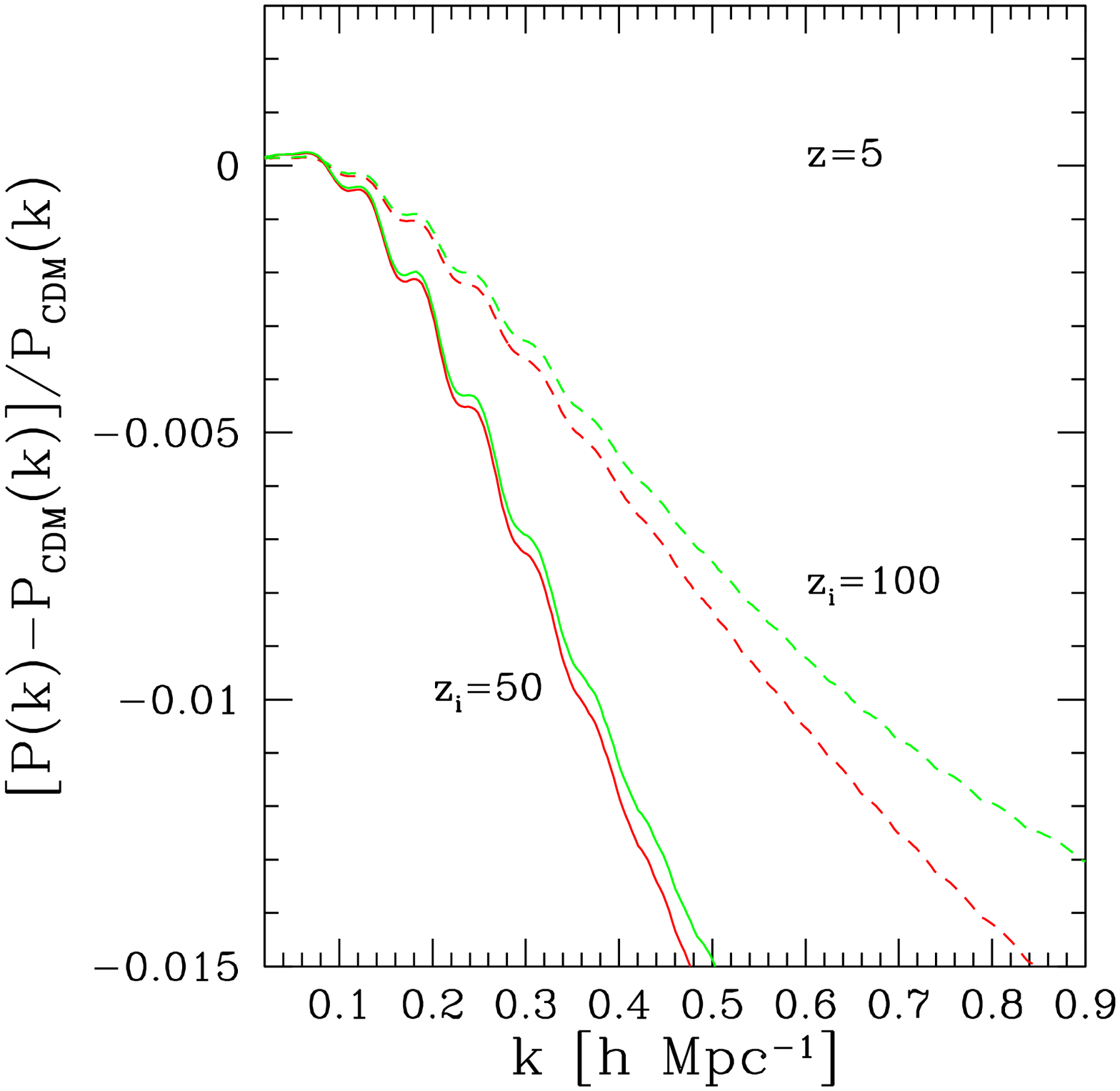}}
\epsfxsize=5.9 cm \epsfysize=5.4 cm {\epsfbox{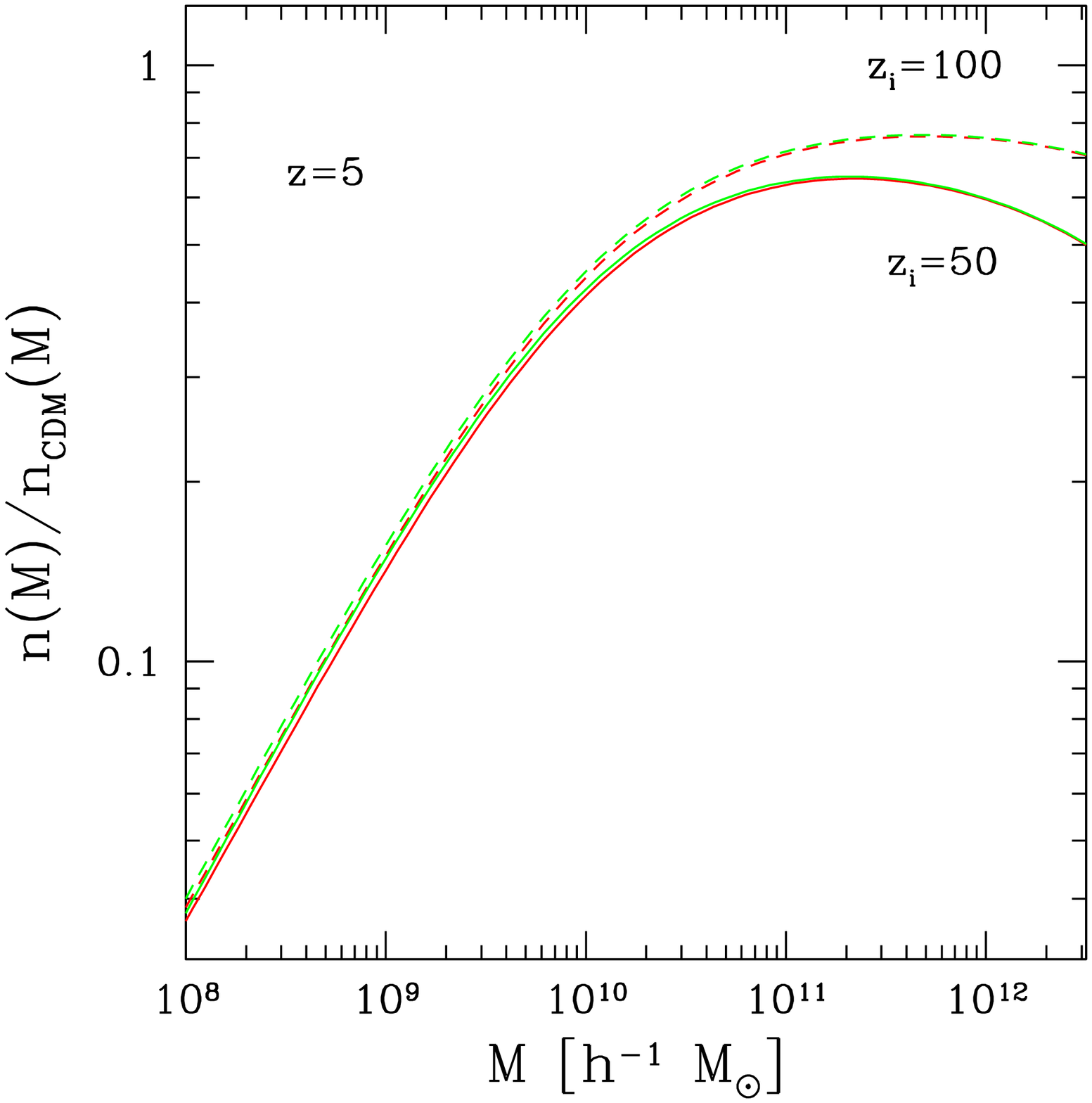}}
\epsfxsize=5.9 cm \epsfysize=5.4 cm {\epsfbox{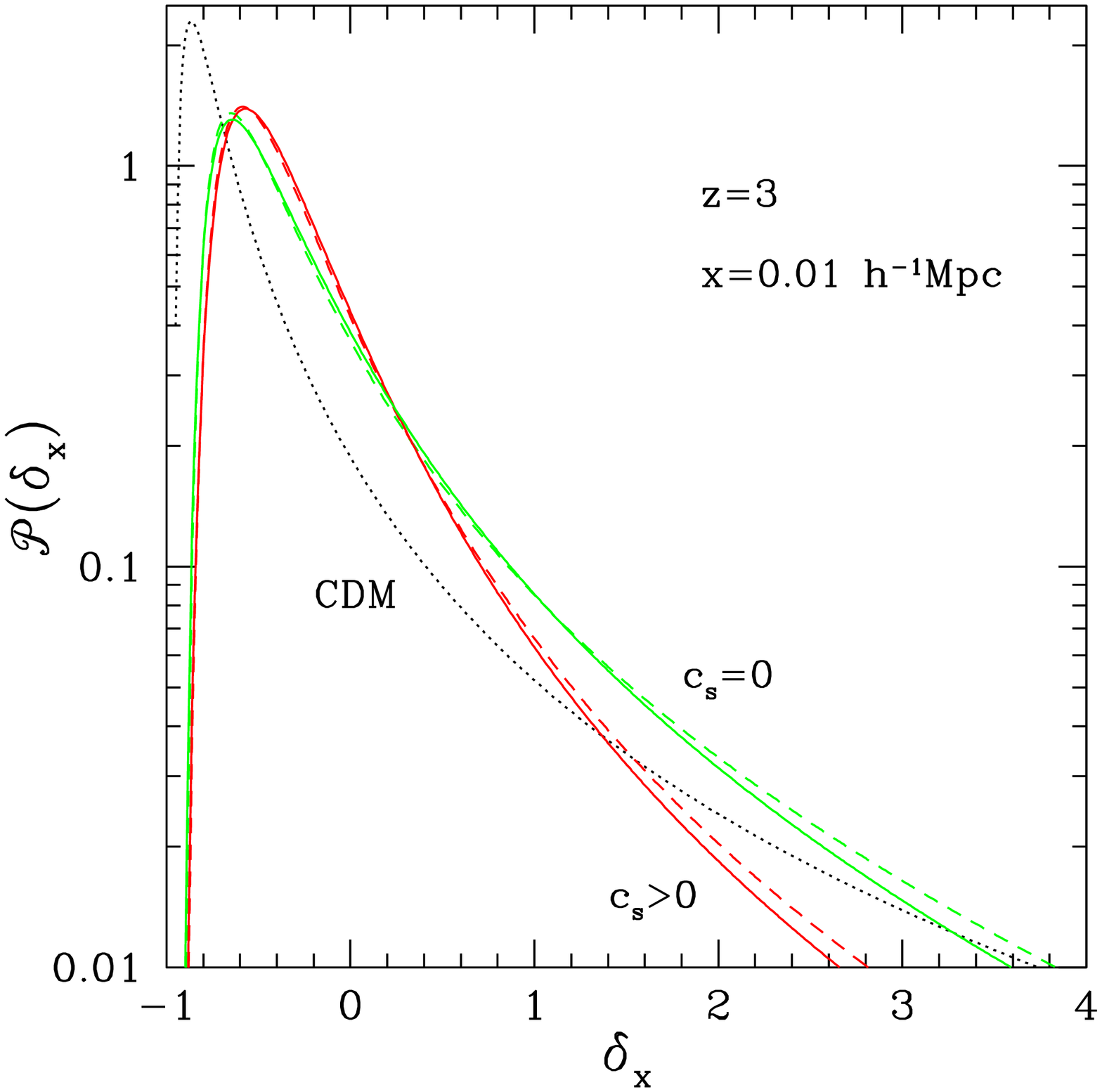}}
\end{center}
\caption{{\it Left panel:} relative deviation of the power spectrum from the CDM reference,
as in Fig.~\ref{fig-dPk}, using an initial redshift $z_i=50$ (solid lines) or $z_i=100$
(dashed lines).
{\it Middle panel:} ratio of the WDM halo mass function to the CDM halo mass function,
with $z_i=50$ (solid lines) or $z_i=100$ (dashed lines).
{\it Right panel:} probability distribution of the matter density contrast, as in
Fig.~\ref{fig-lPrho}, for the CDM case, and for the WDM case with $z_i=50$ (solid lines)
and $z_i=100$ (dashed lines).
In each panel, the green curves (which correspond to slightly higher $P(k)$ and higher
$\cP(\delta_x)$ at $\delta_x > 2$) are obtained for $c_s=0$ and the lower red curves are for
$c_s\neq 0$.}
\label{fig-zi50}
\end{figure*}

Here we briefly comment on the validity of our method for the results described in the
previous sections.
Two effects are not fully included in our treatment of the dynamics:

a) nonlinearities beyond one-loop order for the computation of the power
spectrum,

b) effects associated with the velocity dispersion that go beyond an effective pressure.

The errors associated with a) can be estimated in Fig.~\ref{fig-BAO} by the comparison
between the standard perturbation theory and the resummation scheme.
Even though the latter is systematically more accurate, the difference
between the two results gives an estimate of the error of the theoretical
predictions on these scales.
Then, as already noticed, one can safely conclude that, in this
regime, the effect of the late-time velocity dispersion is on the order
or smaller than the accuracy of standard perturbation theory.

The errors associated with b) can be estimated from the difference
between the cases $c_s=0$ and $c_s>0$. As explained in Sec.~\ref{Effective-Euler},
at late times, the main effect of the velocity dispersion should be well described
by the ``pressure'' term in Eqs.(\ref{delta-Cs}) and (\ref{Euler}).
Then, as long as the deviation between the two cases $c_s=0$ and $c_s>0$ is
small, one expects that we obtain the correct order of magnitude.
In the regime where the deviation would be large (at very high
redshift), higher-order corrections are expected to come into play.
Therefore, our method should be sufficient for our purposes, in the
regimes studied here\footnote{It is not possible to compare with current simulations
because they do not include the effect of the velocity dispersion (apart from the
definition of the initial conditions) and our simple approach with this
effective pressure term is already beyond what is done in simulations.
(The same algorithms are used for CDM and WDM in N-body codes.)}.

\section{Impact of lower initial redshift}
\label{lower-zi}

In the previous figures, we set the initial conditions at redshift $z_i=100$. In this section,
we investigate the impact of using a lower initial redshift, $z_i=50$.
This should make the two WDM results, with $c_s=0$ and $c_s\neq 0$, closer to each
other, since the velocity dispersion decreases with time. However, because it also sets
the nonlinear contributions to zero at $z_i=50$ instead of $z_i=100$, this also 
further underestimates gravitational clustering.

In the standard $\Lambda$CDM scenario, we usually take the limit $z_i\rightarrow\infty$
within analytical approaches because one is only interested in accurate predictions
\cite{Valageas2011d,Valageas2011e,Valageas2011f}.
Moreover, numerical simulations often use second-order Lagrangian perturbation
theory for their initial conditions to decrease the sensitivity to the initial redshift
\cite{Scoccimarro1998}.
However, WDM simulations often use linear theory to set up their initial conditions.
Indeed, a second-order implementation should, in principle, make use of a second-order
analysis of the Vlasov equation at early times, while most works use the results obtained
from the linearized Vlasov equation. On the other hand, the choice of the initial redshift
is not so obvious in WDM scenarios, because a high initial redshift $z_i$ can lead to
spurious effects due to inadequate modeling of the large velocity dispersion
\cite{Colin2008}, whereas a low initial redshift alleviates this problem but can lead to
an underestimate of gravitational clustering.

The dependence on the initial redshift $z_i$ is easily included within our
analytic approach as follows. In the perturbative framework, used for the large-scale
power spectrum shown in Figs.~\ref{fig-Dplin_k}-\ref{fig-BAO}, the integrals over time
associated with the one-loop and higher-order contributions run from $z_i$ down to the
redshift $z$ of interest. In the spherical dynamics, used for the mass function and the
density probability distribution shown in Figs.~\ref{fig-lnM} and \ref{fig-lPrho},
the equations of motion are also integrated from the initial redshift $z_i$.
This allows us to obtain the dependence on $z_i$ of the large-scale structures built at a given
redshift $z$, as in numerical simulations initialized at linear order at this redshift $z_i$.
We show our results for the density power spectrum, the halo mass function, and the
density probability distribution function, comparing the choices $z_i=50$ and
$z_i=100$, in Fig.~\ref{fig-zi50}\footnote{We choose these two values because they
are commonly used in numerical simulations. Moreover,
considering high values such as $z_i=1000$ would not be so useful because simulations
avoid such high initial redshifts to save computational time and to avoid the regime
where the impact of velocity dispersion is large (which is not
included in a rigorous manner in N-body codes).}.

The left panel shows that on perturbative scales, $k<0.9 h$Mpc$^{-1}$ at $z=5$, using an initial
redshift of $z_i=50$ leads to an underestimation of the power spectrum on the order of $1\%$.
Moreover, this is larger than the decrease of the power spectrum due to the change from the
CDM to the WDM power spectrum. On the other hand, the difference between the
$c_s=0$ and $c_s\neq 0$ results is very small.
The same behavior is found for the halo mass function, as shown by the middle panel, with an
underestimation of the large-mass tail because of the low initialization redshift $z_i$
that is on the order of or larger than the true decrease due to the WDM scenario.
The low-mass tail is less affected by the value of the initial redshift since it corresponds
to moderate density fluctuations. In contrast, the large-mass tail corresponds to rare events,
which amplifies the sensitivity to the initial conditions.
This also explains why we find in the right panel that using a low initial redshift $z_i=50$ does not significantly
change the probability distribution of the density contrast at $z=3$, for moderate density
fluctuations. Moreover, the difference between the $c_s=0$ and $c_s\neq 0$ results remains
similar to the one obtained with $z_i=100$.
Therefore, starting at $z_i=50$ instead of $z_i=100$ degrades the accuracy of measures 
of gravitational clustering at low redshifts, since it significantly underestimates the
nonlinearities (as measured by the large-scale power spectrum or the halo mass function) and
contaminates the signal associated with the WDM high-$k$ cutoff (on large perturbative
scales or on the large-mass tail of the halo mass function). 
Moreover, using $z_i=50$ does not help to reduce the effect of the late-time velocity dispersion
on the shape of $\cP(\delta_x)$ on Lyman-$\alpha$ scales.
Thus, numerical simulations should use a high initial redshift, $z_i \geq 100$,
rather than a low value, $z_i \leq 50$.

\section{Conclusion}
\label{Conclusion}

Using an effective Euler equation, that agrees with the Vlasov equation at the linear
level (except for subdominant memory terms), we have estimated the impact of a
late-time WDM velocity dispersion on the formation of large-scale structures.
We have only considered the ``cosmic web'', that is, large perturbative scales, moderate
density fluctuations, and the number counts of virialized halos, which can be studied
with analytic tools.
We have focused on the case of a $1$keV dark matter particle, which is representative
of current WDM scenarios (lower masses are excluded by observations, such as
Lyman-$\alpha$ forest data, while higher masses become indistinguishable from the
CDM limit).

We find that on perturbative scales, the deviation of the density power spectrum
from the CDM case is only on the order of $1\%$, at $z \leq 5$, even though it is
slightly amplified by the nonlinear dynamics. This is below the accuracy of the standard
perturbative expansion and requires efficient perturbative schemes. On the other hand,
the effects of the late-time velocity dispersion are negligible over most of the perturbative
range at $z \leq 5$ (so that one could use the same perturbative approaches devised for the
CDM case).
We also find that the late-time velocity dispersion has a negligible impact on the halo
mass function at $z \leq 5$ (in any case, below the $10\%$ accuracy that can be
guaranteed by simulations), although at very low mass and high redshift, the cutoff
may become sharper \cite{Barkana2001,Benson2012}.
On the other hand, it has a non-negligible effect on the probability distribution of the density
contrast on scales $x \leq 0.1 h^{-1}$Mpc at $z=3$. This means it should have some
impact on the probability distribution of the Lyman-$\alpha$ flux decrement, measured on
the spectra of distant quasars.

Finally, we note that numerical simulations should use a high initial redshift, $z_i \geq 100$,
rather than a low value, $z_i \leq 50$. Indeed, such a low initial redshift can lead to
a significant underestimation of the power spectrum on perturbative scales and of the
large-mass tail of the halo mass function, which is larger than the true signal associated
with the WDM scenario (but of course, on smaller scales and on the low mass tail of the
mass function, one is again dominated by the actual WDM signal).
A low initial redshift does not help much either to reduce the effect of the late-time velocity
dispersion for the probability distribution of the
density contrast on scales associated with Lyman-$\alpha$ clouds.

To go beyond the effective hydrodynamical equations used in this work, one should use
the nonlinear Vlasov equation itself. However, this is a difficult task because of the
additional velocity coordinates, which makes numerical implementations significantly 
heavier already for the CDM scenario \cite{Valageas2004,Tassev2011}.
An alternative would be to extend the fluid approximation to higher orders \cite{Shoji2010},
by including equations of motion for the velocity moments of the Vlasov equation up
to some higher order $n \geq 3$. 
However, because most of the WDM signal arises from nonperturbative scales, such
a task may not be very rewarding, unless one builds methods that can be applied to the
Lyman-$\alpha$ forest clouds, for instance.

\acknowledgments{This work was performed using HPC resources from GENCI-CCRT (Grant 2012-t2012046803).}

\bibliography{ref1}   

\end{document}